\documentclass[12pt,a4paper]{article}
\usepackage[utf8]{inputenc}
\usepackage[english]{babel}
\usepackage{amsmath}
\usepackage{amsfonts}
\usepackage{amssymb}
\usepackage{authblk}
\usepackage{graphicx}
\usepackage{wrapfig}
\usepackage{color}
\usepackage{float}
\usepackage{url}
\usepackage{tikz}
\usepackage{cite}
\usepackage{mciteplus}
\usepackage{dsfont}
\pdfoutput=1
\usepackage{ifpdf}
\usepackage{epsfig}
\usetikzlibrary{arrows,decorations.markings}
\usepackage{hyperref}
\hypersetup{
	colorlinks,linkcolor=black,citecolor=blue,urlcolor=blue}
\usepackage[left=2cm,right=2cm,top=2.6cm,bottom=2.7cm]{geometry}
\usepackage[width=0.8\textwidth,font=small,labelfont=bf]{caption}
\newcommand{\be}{\begin{equation}}
\newcommand{\ee}{\end{equation}}
\newcommand{\beq}{\begin{equation}}
\newcommand{\eeq}{\end{equation}}
\newcommand{\bea}{\begin{eqnarray}}
\newcommand{\eea}{\end{eqnarray}}
\newcommand{\ena}{\end{eqnarray}}

\renewcommand{\a}{\alpha}

\renewcommand{\th}{\theta}

\newcommand{\g}{\gamma}

\newcommand{\e}{\epsilon}

\renewcommand{\k}{\kappa}

\renewcommand{\l}{\lambda}

\newcommand{\vf}{\varphi}

\newcommand{\p}{\pi}

\newcommand{\s}{\sigma}

\renewcommand{\t}{\tau}

\def\tr{\textrm{tr}}

\newcommand{\sn}[2]{\text{sn}\left(\left.\! #1\right| #2\right)}

\newcommand{\cn}[2]{\text{cn}\left(\left.\! #1\right| #2\right)}
\newcommand{\nc}[2]{\text{nc}\left(\left.\! #1\right| #2\right)}
\newcommand{\dn}[2]{\text{dn}\left(\left.\! #1\right| #2\right)}
\newcommand{\nd}[2]{\text{nd}\left(\left.\! #1\right| #2\right)}

\newcommand{\ns}[2]{\text{ns}\left(\left.\! #1\right| #2\right)}

\numberwithin{equation}{section}

\DeclareMathOperator{\sech}{\text{sech}}
\DeclareMathOperator{\am}{\text{am}}

\newcommand{\elF}[2]{\mathds{F}\left(\left.#1\right|#2\right)}
\newcommand{\elE}[2]{\mathds{E}\left(\left.#1\right|#2\right)}
\newcommand{\eE}{\mathds{E}}
\newcommand{\eK}{\mathds{K}}

\newcommand{\eP}[2]{\Pi\left(\left.#1\right|#2\right)}

\title{\textbf{Wilson loops correlators in defect $\mathcal{N}=4$ SYM}}
\author[1,2]{Sara Bonansea\footnote{\href{sara.bonansea@nbi.ku.dk}{sara.bonansea@nbi.ku.dk}}}
\affil[1]{\small\textit{Dipartimento di Fisica, Università di Firenze and INFN Sezione di Firenze, via G. Sansone 1, 50019
Sesto Fiorentino, Italy}}
\affil[2]{\small\textit{Niels Bohr Institute, Copenhagen University,	Blegdamsvej 17, 2100 Copenhagen, Denmark}}
\author[3]{Renato Sánchez\footnote{\href{renato@if.usp.br}{renato@if.usp.br}}}
\affil[3]{\small\textit{Institute of Physics, University of São Paulo, 05508-090 São Paulo, Brazil}}

\date{}

\begin{document}
\clearpage 
\maketitle
\thispagestyle{empty}

\vskip 3cm

\abstract{\noindent We consider the correlator of two concentric circular Wilson loops with equal radii for arbitrary spatial and internal separation at strong coupling within a defect version of $\mathcal{N}=4$ SYM. Compared to the standard Gross-Ooguri phase transition between connected and disconnected minimal surfaces, a more complicated pattern of saddle-points contributes to the two-circles correlator due to the defect's presence. We analyze the transitions between different kinds of minimal surfaces and their dependence on the setting's numerous parameters.}

\newpage

\setcounter{page}{1}
\tableofcontents
\section{Introduction}
\begin{figure}[t]
	\centering
	\includegraphics[width=0.63 \textwidth]{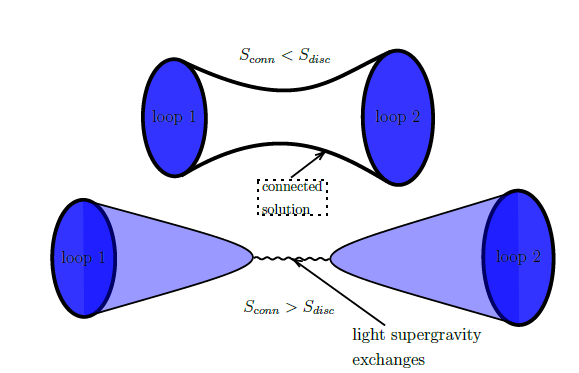}
    \caption{\label{fig1}\footnotesize Different saddle-points for the correlator of two circular Wilson loops. The worldsheet with the topology of an annulus is dominant below a certain critical value for the separation distance between the two circles.}
	\label{plogen1}
\end{figure}

According to the AdS/CFT correspondence, the string theory dual of the Maldacena Wilson loop expectation value in the large 't Hooft coupling limit is described by the area of a minimal surface attached on the contour of the Wilson loop at the conformal boundary of the AdS space \cite{Maldacena:1998im,Rey:1998ik}. The expectation value of the supersymmetric circular Wilson loop, fully captured by a hermitian matrix model in $\mathcal{N}=4$ Super Yang-Mills (SYM) theory \cite{Erickson:2000af,Drukker:2000rr,Pestun:2007rz}, interpolates between strong and weak coupling constituting one of the first positive tests of the AdS/CFT correspondence.
Gross and Ooguri pointed out in  \cite{Gross:1998gk} that in the case of the correlator between two circular Wilson loops a novel feature appears. The minimal surface bounded by the two loops can have the topology of an annulus connecting the two circles or can consist of two separate disk solutions that interact among themselves by the exchange of light supergravity modes \cite{Berenstein:1998ij,Bassetto:2009rt}. The correlator between two circles experiences a sort of phase transition between these two competing saddle-points (see \autoref{fig1}). The two-point function of two concentric circular Wilson loops with the same constant internal space orientation has been studied both for equal \cite{Zarembo:1999bu} or different radii \cite{Olesen:2000ji}. In \cite{Kim:2001td}, the correlator of Wilson loops with the same radii is examined at both zero and finite temperatures. The result is that for large separation between the two loops, the preferred configuration is the one given by separate surfaces. The area of the annulus increases with the separation between the loops and becomes energetically less favorite. In \cite{Correa:2018lyl}, this study further generalizes to concentric loops with both spatial and internal separation. From a perturbative point of view, to analyze this phase transition, in principle, one has to sum all planar diagrams. Thus, if one considers the strong coupling limit of the ladder diagrams resummation does not expect to match the string theory computation, because the interacting diagrams are not taken into account. Nevertheless, it is possible to find a qualitative matching since the ladders resummation presents a phase transition that resembles the Gross-Ooguri one \cite{Zarembo:2001jp,Correa:2018lyl,Correa:2018pfn}.

\begin{figure}[t]
\includegraphics[width=0.95\textwidth]{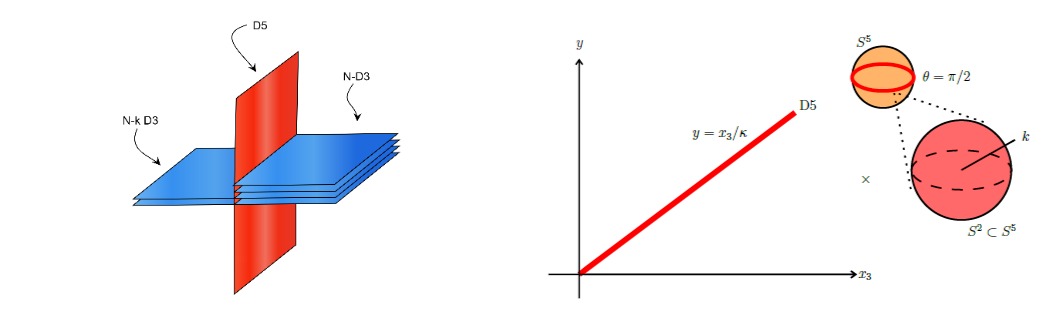}
\caption{\footnotesize{D3-D5 system: $k$ of the D3-branes end on the probe D5-brane which in AdS is tilted with respect to the boundary $y = 0$ at an angle that depends on $\k$. }}
\label{fig2}
\end{figure}

\noindent
An interesting new arena where consider circular Wilson loops and their correlator is a defect version of $\;\mathcal{N}=4$ SYM theory \cite{DeWolfe:2001pq,Erdmenger:2002ex,Nagasaki:2011ue}. The general study of defects is an important subject, which has relations with the physics of mostly every field theory. Spatial defects can be introduced into a conformal field theory (CFT) as means to make contact with the real world, reducing the total amount of symmetry and making the correlation functions less constrained than for a usual CFT. In this paper, we consider a codimension-one defect located at $x_3 = 0$ that separates two regions of space-time where the gauge groups is $SU(N)$ for $x_3>0$ and $SU(N - k)$ for $x_3<0$, see \cite{deLeeuw:2017cop} for a recent review. In the field theory description, the difference in the rank of the gauge groups is achieved by considering a non-trivial vacuum solution in which three of the $\mathcal{N}=4$ SYM scalar fields acquire a non-vanishing vacuum expectation value (VEV) proportional to $1/x_3$ for positive values of the coordinate perpendicular to the defect
\begin{equation}
\left\langle\phi_i \right\rangle_{\text{tree}}= \phi_i^{cl} =-\frac{1}{x_3} t_i \oplus 0_{(N-k)\times(N-k)} \qquad \quad i=1,2,3\quad \text{for} \; x_3>0 \;,
\end{equation}
where $t_i$ are the $\,\mathfrak{su}(2)$ generators in a $k$-dimensional irreducible representation. All the classical fields are vanishing for $x_3<0$. The VEVs are solution to the Nahm's equations \cite{Nahm:1979yw,Diaconescu:1996rk,Constable:1999ac} that arise as conditions for the defect to preserve half of the original supersymmetry \cite{Gaiotto:2008sa}. 
The four-dimensional conformal group $SO(2,4)$ of $\mathcal{N}=4$ SYM is reduced to the three-dimensional one $SO(2,3)$ by the presence of the defect and the R-symmetry group $SO(6)$ is broken down to $SO(3) \times SO(3)$. The full superconformal group preserved by the defect is $OSp(4\left| \right.4)$. The field theory configuration has a holographic dual realized by $N$ D3 branes intersected by a single probe D5 brane which in the near horizon limit warps an $AdS_4 \times S^2$ geometry inside the $AdS_5 \times S^5$ background generated by the D3 branes \cite{Karch:2000ct,Karch:2000gx,Karch:2001cw}.
There is also a background gauge field flux of $k$ units through the $S^2$ sphere, meaning that $k$ of the D3 branes get dissolved into the D5 brane, as shown in \autoref{fig2}. The probe brane has worldvolume coordinates $(x_0,\;x_1,\;x_2,\;y,\;\Omega_{S^2})$, it is tilted with an angle, that depends on $k$, respect to the boundary of $AdS_5$ located at $y=0$
\begin{equation}
y=\frac{x_3}{\k} \qquad  \qquad \text{where} \quad \kappa=\frac{\p k}{\sqrt{\lambda}}\;,
\end{equation}
and sits at the equator of the $S^5$ sphere.
Since the defect breaks part of the conformal symmetry, one-point functions of composite operators can be non-vanishing already at tree-level \cite{Cardy:1984bb}, as analyzed in \cite{Nagasaki:2012re,Kristjansen:2012tn} for chiral primary operators in the defect set-up with $k$ units of flux. Furthermore, two-point functions of operators with unequal scaling dimension can be non-vanishing \cite{deLeeuw:2017dkd}. Using the tool of integrability, it was possible to derive  a closed formula for tree-level one-point functions of non-protected operators in the $SU(2)$ sub-sector of $\mathcal{N}=4$ SYM for any $k$ \cite{deLeeuw:2015hxa,Buhl-Mortensen:2015gfd}, then generalized to higher loop orders \cite{Buhl-Mortensen:2017ind} and to extend it to the $SU(3)$ sector \cite{deLeeuw:2016umh} as well as the full $SO(6)$ scalar sector \cite{deLeeuw:2018mkd}, see \cite{deLeeuw:2019usb} for a review. One-point functions of composite operators have been recently analyzed in the $k=0$ and $k>0$ cases using the supersymmetric localization technique in \cite{Wang:2020seq,Komatsu:2020sup}.   
Compared to the standard AdS/CFT scenario, the defect that we are considering is characterized by the presence of a novel parameter $k$, that on the field theory side controls the vevs of the scalar fields. In \cite{Nagasaki:2011ue,Nagasaki:2012re}, was suggested to consider a certain double scaling limit which consists of letting $N \rightarrow \infty$ and subsequently sending $k$ and $\lambda$, the 't Hooft coupling constant, to infinity (but with $k<<N$) in such a way that the ratio $\lambda/k^2$ can be taken fixed and small. Since $\lambda \rightarrow \infty$, the supergravity approximation is reliable and for certain observables the results on the field theory and the string theory side of the correspondence organize in power series of $\lambda/k^2$, enabling the gauge-gravity comparison. Positive tests of the AdS/CFT correspondence in the D3-D5 defect set-up where both supersymmetry and conformal symmetry are partially broken have been performed in the double scaling limit for one-point functions of chiral primaries both at tree-level and one-loop\cite{Buhl-Mortensen:2016pxs,Buhl-Mortensen:2016jqo}.
A highly non-trivial matching between gauge and string results for these operators has been achieved also in a different domain-wall version of $\mathcal{N}=4$ described by a D3-D7 probe system, where all the supersymmetries are broken \cite{Kristjansen:2012tn,deLeeuw:2016ofj,Grau:2018keb,Gimenez-Grau:2019fld,deLeeuw:2019ebw}, see \cite{Linardopoulos:2020jck} for a recent review. Moreover, an agreement between gauge and string theory calculations in the D3-D5 set-up employing the double
scaling limit was found for a single Wilson line \cite{Nagasaki:2011ue,deLeeuw:2016vgp}, a pair of Wilson lines \cite{Preti:2017fhw}, and a circular Wilson loop in \cite{Aguilera-Damia:2016bqv}, for a small value of the coupling $\chi$ with the scalars and small distance from the defect, and in \cite{Bonansea:2019rxh} where the double scaling limit was performed without restrictions on $\chi$ and on the distance from the interface. A positive check in the D3-D7 system has been performed for the Wilson line in \cite{Bonansea:2020vqq}.
 Focusing on a circular Wilson loop of radius $R$ placed at $x_3=L$ on a plane parallel to the defect in the D3-D5 brane system, it has been shown in \cite{Bonansea:2019rxh} that it experiences a first-order phase transition of Gross-Ooguri type (GO-like). The disk solution, which is the minimal surface in the theory without the defect, dominates when the operator is far from the interface. On the other hand, the cylindrical string solution, connecting the boundary loop with the probe D5-brane, is favorite below a certain critical distance from the defect. \\ \\
\noindent
In this work, we study the gravity dual of the two concentric circular Wilson loops correlator in the defect version of $\mathcal{N}=4$ SYM theory. We will handle a complicated pattern of saddle-point configurations that are dominant in different regions of the parameter space describing the solutions. Besides the two saddle-points given by the disk-disk or the connected configuration between the circles, the minimal area configuration in the defect set-up can be given by two connected cylindrical surfaces attached to the boundary of AdS and ending on the D5 brane, or we can also have an intermediate case with a cylindrical surface for one of the loops and a disk for the other one. In the single circle case, the cylindrical surface attached to the defect can be described in terms of three parameters: $\k$, the coupling $\chi$ of the loop with the scalars, and $L/R$. Since the defect preserves a subgroup of the original four-dimensional conformal group, the Wilson loop expectation value will depend on $R$ and $L$ only through their ratio. If two circles are considered, in addition to $\k$ we can have two different couplings with the scalars, $\chi_1$ and $\chi_2$, two different radii for the circles, $R_1$ and $R_2$, and two different positions along $x_3$  that we indicate as $L_1 \equiv L$ and $L_2 = L+h$, where $h$ is the separation between the two loops. Thus, the parameter space describing the different saddle-points is enlarged. To simplify the problem, we will focus on the equal radii case. We will study the different sets of transitions \`a la Gross-Ooguri between different types of minimal surfaces and how they depend on the numerous parameters of the setting. This paper is organized as follows. In sec. \ref{sec2} we find a new way to parametrize the connected solution between two circles and we analyze the allowed region of variation for the new parameters. This will make easier the comparison with the connected worldsheet solution to the defect, given in \cite{Bonansea:2019rxh}. In sec. \ref{sec3}, we describe the feasible pattern of phase transitions occurring when we vary the parameters, focusing on some illustrative scenarios. At the end of this section, we investigate the possibility that the connected solution between the loops is intersected by the defect. A series of appendices complete the paper.

\section{Connected solution between two circles}
\label{sec2}
\subsection{New parametrization for the ``undefected" solution}
The parametrization for the connected string solution between two concentric circles with different coupling with the scalars and radii has been carried out in \cite{Correa:2018lyl}. Here, the parametrization of the relevant quantities as the renormalized action $S_{\text{ren}}$, the angular separation $\gamma$, and the distance between the two circles $h$, is given in terms of two independent parameters $s$ and $t$ that have to be real and satisfy the condition $s \leq 1-t$. In the presence of the defect, another saddle-point, given by two separate cylindrical surfaces that start from the boundary of AdS and attach to the D5-brane, contributes to the evaluation of the correlator between the two circles. The exact solution for the cylindrical minimal area was found in \cite{Bonansea:2019rxh}. In the following, it is more convenient to reparametrize the standard connected solution between the loops to simplify the comparison with the cylindrical surfaces, that will be the object of the analysis in sec \ref{sec3}. The Euclidean metric of $AdS_5 \times S^5$ is parametrized  in Poincar\`e  coordinates as
\be
ds^2=\frac{1}{y^2}(d t^2 +dy^2+dr^2+r^2 d\phi^2+dx_3^2)+d\theta^2+\sin^2 \theta d \Omega^2_{(1)}+\cos^2\theta  d\Omega^2_{(2)},
\ee
where $d\Omega^2_{(i)}=d\alpha_i^2+\sin^2\alpha_id\beta_i^2$ denotes the metric of a 2-spheres in the $S^5$. If we consider the ansatz for the minimal surface used in \cite{Correa:2018lyl,Zarembo:1999bu}
\be
\label{ansatz}
y=y(\sigma),\quad \quad r=r(\sigma),\quad\quad \phi=\tau,\quad\quad x_3=x_3(\sigma)\quad\mathrm{and}\quad \theta=\theta(\sigma),
\ee
we immediately recognize that it is the same string embedding used for the connected solution to the defect. In particular, we can find the minimal area between the two circles taking $x_3$ as a generic function of $\sigma$, without using the gauge choice $x_3=\sigma$. The exact geometric solution for the minimal surfaces described by this ansatz has been found in \cite{Bonansea:2019rxh}, and its embedding in $AdS_5\times S^5$ is given by  
\begin{align}
\label{solution}
AdS_5:\quad
y(\sigma)&=\frac{ R\cosh\eta}{\sqrt{1+g^2(\sigma)}} \textrm{sech} [v(\sigma)-\eta]\quad\quad
r(\sigma)=R\cosh\eta \frac{g(\sigma)}{\sqrt{1+g^2(\sigma)}} \textrm{sech} [v(\sigma)-\eta]\nonumber \\
x_3(\sigma)&= L+R\sinh \eta+R\cosh\eta\tanh [v(\sigma)-\eta]\nonumber\\
S^5: \quad \theta(\s)&=j\s+l\;,
\end{align}
where $\s$ is the spatial worldsheet parameter, $\eta$, $j$ and $l$ are integration constants and $g(\s)=\frac{r(\s)}{y(\s)}$ is an auxiliary function. Following \cite{Bonansea:2019rxh}, we can define $v'(\s)$ in terms of $g(\s)$ as 
\be
v'(\s)=\frac{\sqrt{\e_0}}{g^2(\s)+1}\,
\ee
with
\be
g(\s)=\sqrt{n}\,\ns{\sqrt{n}\s}{m},
\ee
such that
\be
v(\s)=\sqrt{\frac{\epsilon_0}{n}}\left[\mathds{F}(\left.\vf\right|m)-\Pi\left(\left.-\frac{1}{n},\vf \right|m\right)\right],\quad\vf=\am\left(\sqrt{n}\s\right),
\ee
where $\vf$ is the Jacobi amplitude, $\mathds{F}(\left.\vf\right|m)$ and $\Pi\left(\left.-\frac{1}{n},\vf\right|m\right)$ are the incomplete elliptic integral of first and third kind with elliptic modulus $m$, respectively, defined in \autoref{elliptics}. Here, $n$ is defined as
\be
n \equiv \frac{j^2-1}{m+1}\;,
\ee
and 
\be
\label{eps0}
\e_0=-\frac{(j^2+m)(j^2 m+1)}{(m+1)^2}
\ee
 is a positive integration constant that appears in the first integral of $g(\s)$. The constraint $\e_0\geq0$ leads to two different ranges of variation for the couple of parameters $(m,j^2)$ :
\be
\label{region}
\text{(a)}:\quad  -1\leq m \leq 0 \quad \text{and}\quad j^2\geq -\frac{1}{m} \qquad \text{(b)}:\quad m<-1 \quad \text{and} \quad j^2 \leq -\frac{1}{m}\,.
\ee
Notice that $y(\s)$, $r(\s)$ and  $x_3(\s)$ in \eqref{solution} satisfy the boundary conditions 
\be
r(0)=R\,, \qquad y(0)=0\,, \qquad x_3(0)=L\,.
\ee
 The parametric solution in \eqref{solution} draws a sub-manifold inside the following $S^3$ in $AdS_5$:
\be
\begin{split}
	(x_3-R \sinh \eta-L)^2+y^2+r^2=&R^2\cosh^2\eta\Big[\frac{ \textrm{sech}^2 [v(\sigma)-\eta]}{1+g^2(\sigma)} +
	\frac{g^2(\sigma)}{1+g^2(\sigma)} \textrm{sech}^2 [v(\sigma)-\eta]+\\&\hspace{2cm}+\tanh^2 [v(\sigma)-\eta]\Big]=R^2\cosh^2\eta\,.
\end{split}
\ee
This surface intersects the boundary of $AdS_5$ at $\sigma=0$, reaching it again at
\be
\label{sigmah}
\hat\sigma=\frac{2}{\sqrt{n}}\mathds{K}(m). 
\ee
In fact, for this value of the spatial world-sheet parameter
\be
g(\hat{\s}) \rightarrow \infty \quad \Rightarrow \quad y(\hat{\s}) \rightarrow 0
\ee
with
\be
\label{v1}
v(\hat\sigma)=2\sqrt{\frac{\epsilon_0}{n}}\left[ \mathds{K}(m)-\Pi\left(\left.-\frac{1}{n}\right|m\right)\right],
\ee
where $\mathds{K}(m)$ and $\Pi\left(\left.-\frac{1}{n}\right|m\right)$ are the complete elliptic integrals of first and third kind. 
We still have to impose the following Dirichlet boundary conditions in $\hat{\s}$
\be
\label{bchat}
r(\hat{\s})=R, \qquad \quad x_3(\hat{\s})=L+h\,,
\ee
being $h$ the separation between the two circles.
Notice that with respect to the solution given in \cite{Bonansea:2019rxh}, we are considering $x_3$ as an increasing function of $\s$ $(x'_3(\s)>0)$. Thus, the solution for $x_3$ has a different sign here. Requiring to have at $\hat{\s}$ a circle of radius $R$, we find 
\be
\label{eta}
R=R\cosh\eta~ \textrm{sech} \left(v(\hat{\s})-\eta\right) \quad \Rightarrow \quad \eta=\frac{v(\hat{\s})}{2}.
\ee
Imposing the last boundary condition in $\hat{\s}$
\be
x_3(\hat\sigma)=
L+2R\sinh\left(\frac{v(\hat{\s})}{2}\right)=L+h \,,
\ee
one finds
\be
\label{h1}
\frac{h}{R}=2\sinh \left(\frac{v(\hat{\s})}{2}\right) \,.
\ee
The boundary condition in $\s=0$ for the angle $\theta$ on the $S^5$ determines the constant $l$
\be
\theta(0)=\chi_1 \quad \Rightarrow \quad l=\chi_1\,.
\ee
In $\hat{\s}$ we have to impose
 \be
  \theta(\hat{\s})=\chi_2=j\hat{\s}+\chi_1\,,
\ee
where $\chi_1$ and $\chi_2$ are the different values of the scalar coupling at the boundary of $AdS_5$ for the two circles.
We can define the angular separation $\gamma$ between the loops as
\be
\label{gamma}
\gamma=\left| \Delta \chi\right|  = \left| \chi_2-\chi_1\right| =\left| j\right|\hat{\s}=\frac{2\left| j\right| }{\sqrt{n}}\mathds{K}(m)\,,
\ee
where $\gamma \in [0,\pi]$. Due to the $SO(6)$ R-symmetry, only the the angular separation $\gamma$ is important and not the individual values of the couplings with the scalars.
 \subsection{Allowed region for the parameters}
\begin{figure}[t!]
	\centering
	\includegraphics[width=.8
	\textwidth]{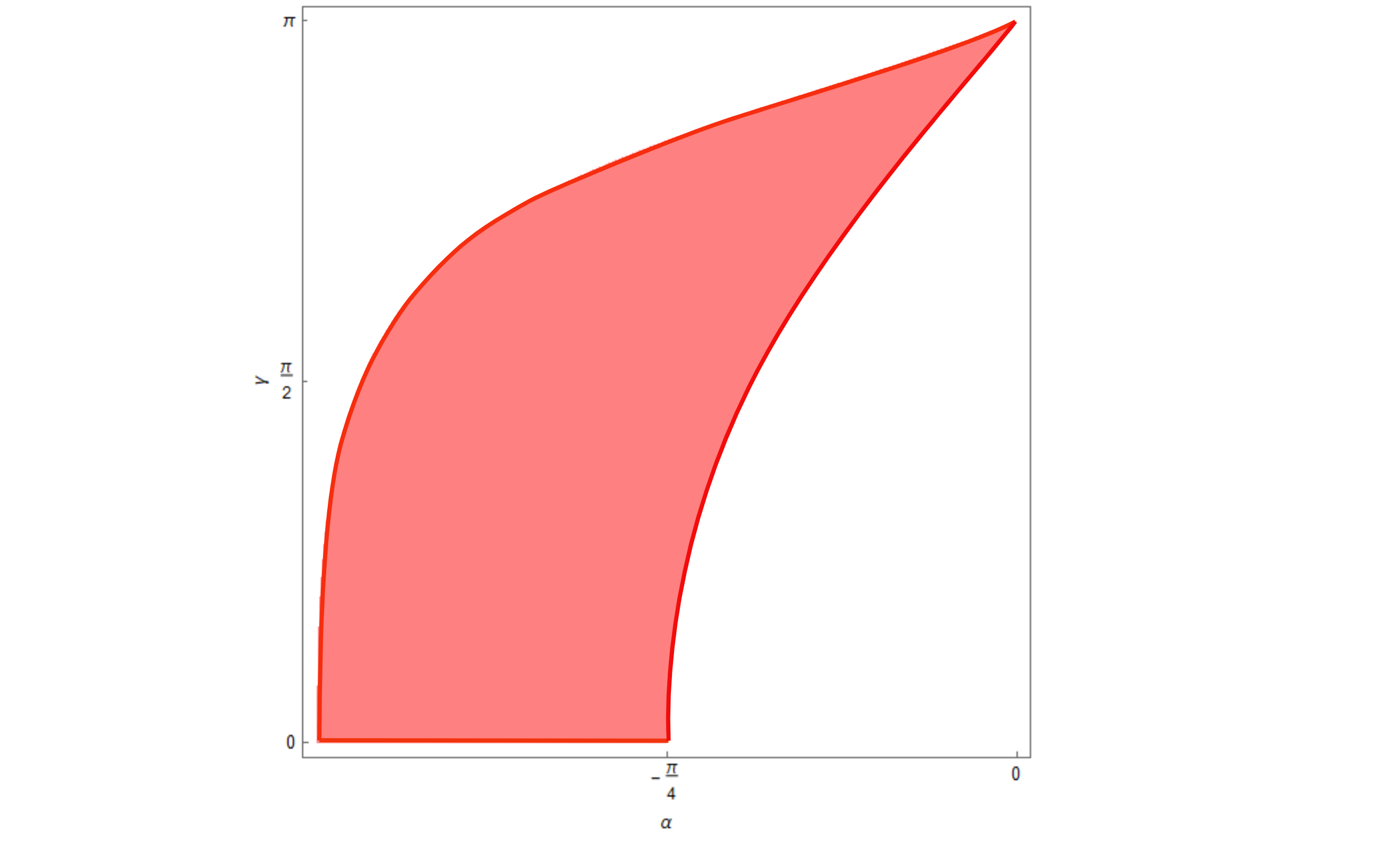}
	\caption{\label{fig3allowed}\footnotesize{Allowed region of parameters in the $(\a,\g)$-plane, where $\a \equiv \arctan m$.}}
\end{figure}

To have a more compact notation, it is possible to introduce the auxiliary parameter 
\be
\label{defx}
x=\sqrt{\frac{j^2-1}{j^2(m+1)}}\,.
\ee
By construction $x$ has to satisfy the following constraints
\be
\label{boundsx}
\text{(a)}: \quad 1 \leq x \leq  \frac{1}{\sqrt{m+1}} \qquad \text{(b)}: \quad x\geq 1\,.
\ee
One can use \eqref{gamma} to determine $x$ in terms of $m$ and $\gamma$
\be
\label{eq:2.22}
x=\frac{2 }{\gamma} \mathds{K}(m).
\ee
The separation distance $h$ can be written as a functions of the angle $\gamma$ and of the modular parameter $m$ using \eqref{h1} and \eqref{v1}
\begin{align}
\label{separation}
&\frac{h}{R}=2 \sinh \left(\sqrt{\frac{\left(x^2-1\right) \left(m x^2-1\right)}{x^2 \left((m+1) x^2-1\right)}} \left(\mathds{K}(m)-\Pi
\left(m-\frac{1}{x^2}+1\right|m\right)\right)=\\
&=2 \sinh \left[\frac{1}{2} \sqrt{\frac{\left(\gamma ^2-4
		\mathds{K}(m)^2\right) \left(\gamma ^2-4 m 	\mathds{K}(m)^2\right)}{4 (m+1)
			\mathds{K}(m)^4-\gamma ^2 	\mathds{K}(m)^2}} \left(	\mathds{K}(m)-\Pi
\left(\left.-\frac{\gamma ^2}{4
		\mathds{K}(m)^2}+m+1\right|m\right)\right)\right]\nonumber\,.
\end{align}
The bounds on $x$ given in \eqref{boundsx}, precisely ensure the positivity of the expression in the square root in \eqref{separation}, and determine an upper and a lower limiting values for $m$ that we denote as $m_a$ and $m_b$, respectively.
In particular, $m_a$ solves the equation
\be
x=\frac{1}{\sqrt{1+m_a}}=\frac{2}{\gamma}\,\mathds{K}{(m_a)}\;,
\ee
while $m_b$ is a solution of
\be
x=\frac{2}{\gamma}\,\mathds{K}(m_b)=1\;.
\ee
Fixing $\gamma$, one can obtain the corresponding values for $m_a$ and $m_b$ such that no solution exists for $m>m_a$ or $m<m_b$. These bounds exposed above depict an allowed region of parameters that we show in \autoref{fig3allowed}.
\vspace{0.5cm}\\
Now, we want to analyze how the separation distance $h/R$ behaves as $m$ approaches the two extrema $m_a$ and $m_b$. 
\noindent
When $m \rightarrow m_a$, we can use the following expansion for $x$
\be
\label{expxa}
x=\frac{1}{\sqrt{1+m_a}}+A(m-m_a)+O((m-m_a)^2)\,,
\ee 
where $A$ is a positive coefficient. We can use the expression of $h/R$ in terms of $x$ and $m$ given in the first line of \eqref{separation}, expanding around $m_a$ we find
\be
\label{limith}
\frac{h}{R}=2\sqrt{-\frac{1+2A(1+m_a)^{3/2}}{m_a}}\,\left( \mathds{E}(m_a)-\mathds{K}(m_a)\right)\sqrt{m_a-m}+O\left((m_a-m) ^{3/2}\right)\;.
\ee
Thus, we can conclude that the separation distance goes to zero as $m$ approaches $m_a$.  Notice that the coefficient $A$ in \eqref{expxa} can be determined by solving \eqref{eq:2.22}  perturbatively, getting
\be
A=\frac{\mathds{E}(m_a)+(m_a-1)\mathds{K}(m_a)}{2(1-m_a)m_a\sqrt{1+m_a}\mathds{K}(m_a)}\,.
\ee
\vspace{0.3cm}\\
When $m \rightarrow m_b$, $x$ goes to $1$ and it is easy to see that $h/R$ goes to zero since the coefficient in the square root of \eqref{separation} vanishes while
\be
\Pi(m \left. \right| m)=\frac{\mathds{E}(m)}{1-m}\,.
\ee
Thus, the distance between the loops is zero when $m$ approaches $m_a$, it increases as $m$ is lowered, it reaches a maximum and it starts to decrease to be again zero at $m=m_b$. From this analysis, it is clear that $h/R$ cannot be a monotonic function of $m$, meaning that we cannot invert \eqref{separation} to get $m$ as a function of $h/R$ and $\gamma$. Therefore, the two independent parameters that we use to describe the connected solution are $m$ and $\gamma$. 
\begin{figure}[t]
	\centering
	\includegraphics[width=.6
	\textwidth]{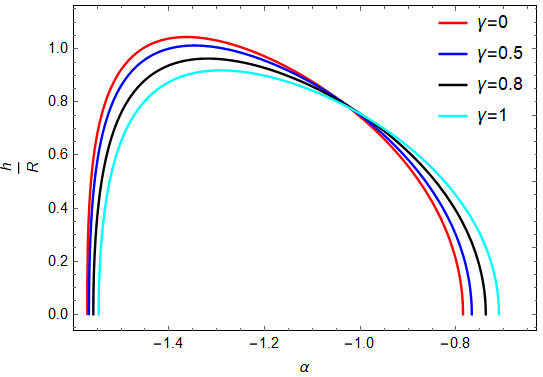}
	\caption{\label{fig4}\footnotesize{The plot displays the behavior of the separation distance between the circles $h/R$ as a function of $\a \equiv \arctan m$. As $\gamma$ increases, $m_a$ becomes larger and closer to 0. For each value of the angular separation, $h/R$ displays a maximum and is not a monotonic function of $m$.}}
\end{figure}
In \autoref{fig4}, is displayed the behavior of $h/R$ as a function of $\a \equiv \arctan m$. For each value of $\gamma$, there is a maximal distance at which the connected solution starts to exist and it becomes smaller as we increase the value of $\gamma$. This result is in agreement with Figure 2 in \cite{Correa:2018lyl}.
\subsection{The area of the connected solution}
\label{2.3}
The action of the connected minimal surface between the two circles is
\be
S=\frac{\sqrt{\lambda}}{2\pi}\int_{\s_{\e}}^{\hat\s_{\e}}d\s d\t \frac{r^2(\s)}{y^2(\s)}=\sqrt{\lambda}\int_{\s_{\e}}^{\hat\s_{\e}}d\s g^2(\s)\,.
\ee
 Since it diverges as the world-sheet approaches the boundary, we have to regularize it imposing the boundary conditions $y(\s_{\e}) =\e$ and $y(\hat\s_{\e})=\e$. Expanding for small $\e$, we have
\be
 \hat\s_{\e}=\frac{2}{\sqrt{n}}\mathds{K}(m)-\frac{\e}{R}-\frac{3+n\,(1+m)}{6\,R^3}\, \e^3+\mathcal{O}(\e^4)\;,
\ee
\be
\s_{\e}=\frac{\e}{R}+\frac{1}{6}(j^2+2)\frac{\e^3}{R^3}+\mathcal{O}(\e^4)\,.
\ee
Thus, the action becomes
\be
S=\sqrt{\l}\left( -2\,\sqrt{n}\left(\mathds{E}(m)-\mathds{K}(m) \right) +\frac{2R}{\e}+\mathcal{O}(\e)\right) \,.
\ee
Removing the divergent piece proportional to the perimeter of the two circles, the expression for the renormalized action takes the form 
\be
\label{action}
\hat{S}_{\text{ren}}=\frac{S_{\text{ren}}}{\sqrt{\lambda}}=-2\,\sqrt{n}\left(\mathds{E}(m)-\mathds{K}(m) \right)=\frac{4 \mathds{K}(m) (\mathds{K}(m)-\mathds{E}(m))}{\sqrt{\gamma ^2-4 (m+1) \mathds{K}(m)^2}}\,.
\ee
\begin{figure}[t!]
	\centering
	\includegraphics[width=.8
	\textwidth]{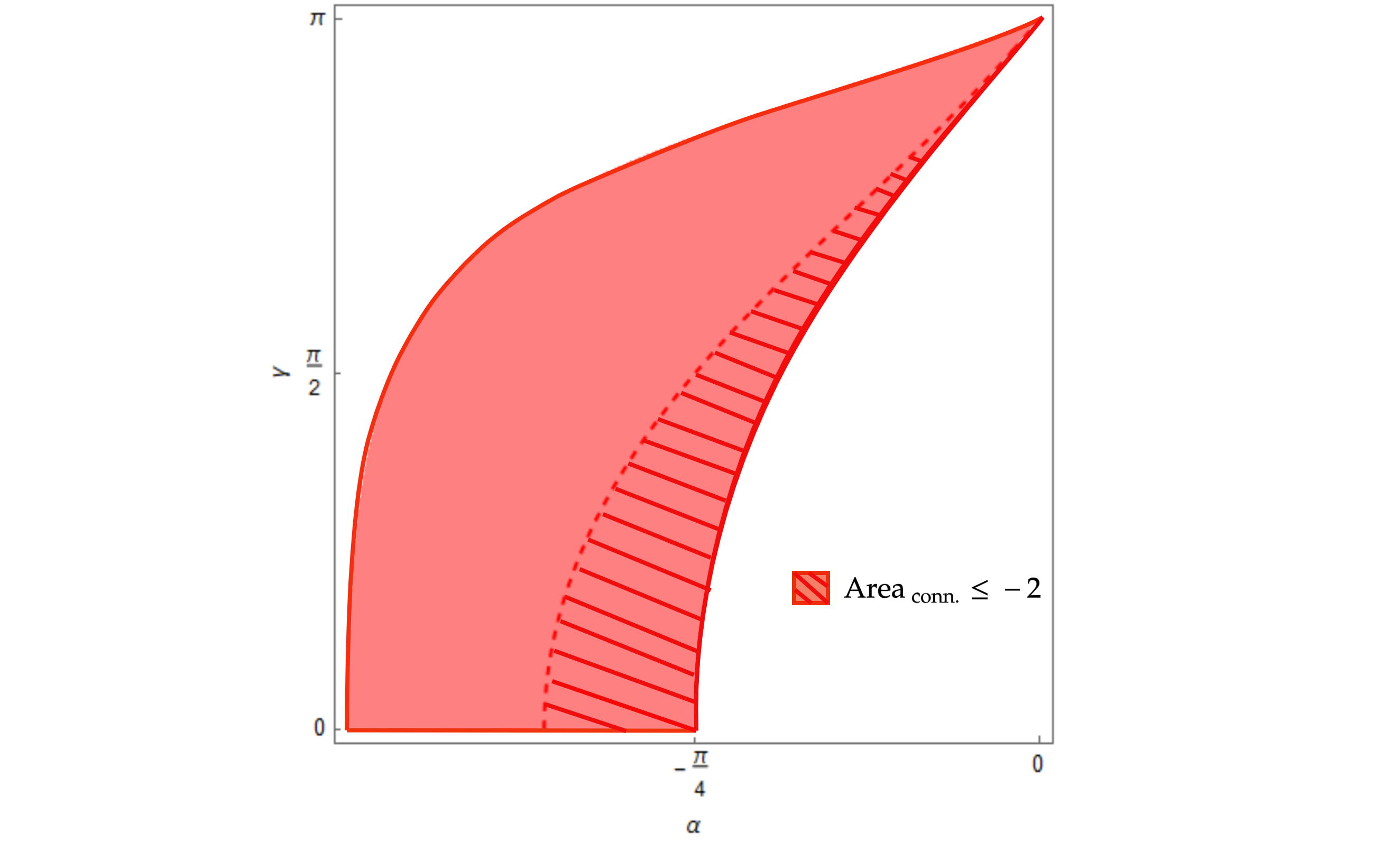}
	\caption{\footnotesize{The region on the right of the dashed red line in the $\a-\g$ plane corresponds to $\hat{S}_{\text{ren}}\leq-2$.}}
	\label{allowed2}
\end{figure}
All the relevant quantities can be parametrized in terms of $m$ and $\gamma$.
It is interesting to analyze the behavior of the area for the two boundary values of $m$. Thus, it is more convenient to rewrite the $\hat{S}_{\text{ren}}$ as a function of $m$ and $x$
\be
\hat{S}_{\text{ren}}=\frac{2(\mathds{K}(m)-\mathds{E}(m))}{\sqrt{\left( \frac{1}{x^2}-(m+1)\right)}}\,.
\ee
When $m \rightarrow m_a$, using \eqref{expxa}, we find 
\be
\hat{S}_{\text{ren}}=\frac{2(\mathds{K}(m_a)-\mathds{E}(m_a))}{ \sqrt{2 A (m_a+1)^{3/2}+1 }}\frac{1}{\sqrt{m_a-m}}+O\left((m_a-m)^{1/2} \right) \,.
\ee
Since the combination $(\mathds{K}(m)-\mathds{E}(m))$ is always negative for $m<0$, we can conclude that the renormalized action diverges to $-\infty$ as $m$ approaches $m_a$. We can get the short-distance asymptotics of the action using the expansion for $h/R$ given in \eqref{limith}
\be
S_{\text{ren}}=-\frac{4(\mathds{K}(m_a)-\mathds{E}(m_a))^2}{\sqrt{-m_a}}\frac{R}{h}\,.
\ee
It exhibits a Coulombic behavior and diverges when the two circles coincide. 
 If $m_a=-1$, which corresponds to the $\g=0$ case, we recover the same result found in eq. (2.32) of \cite{Zarembo:1999bu} (notice that there is a difference of a $2\pi$ factor coming from how we normalized the action).
\\\\ If $m \rightarrow m_b$, the renormalized action assumes the finite value
\be
\hat{S}_{\text{ren}}=2\frac{\mathds{K}(m_b)-\mathds{E}(m_b)}{\sqrt{-m_b}}\,.
\ee
In \autoref{allowed2}, we have plotted the allowed region of parameters, given by the conditions in \eqref{boundsx}, in the $\a-\g$ plane corresponding to $\hat{S}_{\text{ren}}\leq-2$. This region collapses to a point when $\gamma=\pi$, meaning that no connected solution between the two circles exists for this value of the angle. Moreover, $\gamma=\pi$ is a solution for the equation \cite{Correa:2018lyl}
\be
\cos\g=-1-\frac{h^2}{2R^2}
\ee
that corresponds to the condition for the two Wilson loops to have common supersymmetries. In this case, we have two separate domes that interact by the exchange of supergravity fields \cite{Giombi:2009ms}. \\ \\
The area of the connected solution as a function of the separation distance $h/R$ for different values of the angular separation $\gamma$ is presented in \autoref{fig5}. For each $\gamma$, there is always a maximal distance at which the connected solution starts to exist. If we decrease $h/R$ from its maximum, we find two different branches of solutions that terminate at $h/R=0$. The upper one is always subdominant and its end-point at $h/R=0$ corresponds to $m=m_b.$ The lower and dominant branch is given by values of $m$ that run from $m_a$ to the maximal value of the separation.
\begin{figure}[t!]
	\centering
	\includegraphics[width=.6
	\textwidth]{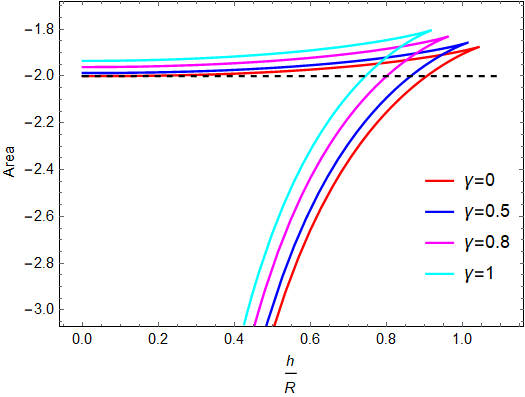}
	\caption{\label{fig5} \footnotesize{The plot displays the behavior of the area of the connected solution as a function of the distance $h/R$ between the two circles with the same radius $R$. When $\gamma$ increases, the value of $h/R$ at which the connected solution starts to exist becomes smaller, and the GO transition occurs at lower separation distances. We have normalized to -2 the area of two spherical domes.}}
\end{figure}
There is always a critical $h/R$ at which the renormalized action $\hat{S}_{\text{ren}}$ is equal to the area of two disconnected spherical domes. Thus, for values of $h/R$ greater than the critical one, the dominant solution is given by two spherical domes. Below the critical separation distance, the configuration with the minimal area is the connected solution. We can numerically determine both the critical and the maximal value of $h/R$. For $\gamma=0$, we have verified that they coincide with the results given in \cite{Zarembo:1999bu}, namely
\be
\left( \frac{h}{R} \right)_{\text{crit}}=0.91 \qquad \left( \frac{h}{R} \right)_{\text{max}}=1.045\,.
\ee
\section{Phase transitions in the presence of the defect}
\label{sec3}
The study of a single circular Wilson loop in $\mathcal{N}=4$ SYM theory in the presence of a defect, considered in \cite{Aguilera-Damia:2016bqv,Bonansea:2019rxh}, can be extended to the case in which two circles are inserted in the set-up realized by the D3-D5 probe-brane intersection. Besides the two different saddle points described in sec. \ref{2.3}, the minimal area configuration can be given by two connected cylindrical surfaces stretched from the boundary of AdS to the defect or a cylindrical surface for one of the loops and a disk for the other one. The two concentric loops are placed on planes parallel to the defect at a distance $x_3=L$ and $x_3=L+h$, respectively. For simplicity, we consider the case in which the circles have the same radius $R$, namely $R_1=R_2=R$. The two Wilson loops can couple with the scalars $\phi_3$ and $\phi_6$ of the theory with different angles and they are opposite oriented in space-time\footnote{No connected solution between the circles is allowed if they have the same orientation \cite{Zarembo:2001jp,Correa:2018lyl,Correa:2018pfn}}
\be
\mathcal{W}\left( C_j\right) =\tr\mathrm{\,Pexp}\left(i\oint (A_\mu\dot x^\mu_j+i |\dot x_j|(\phi_3\sin\chi_j+\phi_6\cos\chi_j))\right) \qquad j=1,2
\ee
where each loop is parametrized by
\begin{align}
	x^\mu_1&=\left(0,\;R\cos\t_1,\;R\sin\t_1,\;L\right), \noindent \\
		x^\mu_2&=\left(0,\;R\cos\t_2,\;-R\sin\t_2,\;L+h \right), \qquad \t_j \in \left[0,\;2\pi \right] \;.
\end{align}
The first loop, located at $x_3=L$, will have an internal angle $\chi_1$ while for the second one, at $x_3=L+h$, the coupling with the scalar will be given by $\chi_2$. The parameters that describe the solutions are
\be
\label{parameters}
\kappa, \quad \chi_1, \quad \chi_2, \quad \frac{L}{R} \quad \text{and} \quad \frac{h}{R}\,.
\ee
As anticipated, when the defect is present we have richer situations with respect to the standard Gross-Ooguri phase transition between the two circles. We expect that when both circles are far away from the defect $(L/R >>1)$, the picture is the one described in the previous section. Depending on $h/R$, the dominant solution will be given by the connected surface between the contours or it will consist of two disconnected domes. If we get closer to the defect, we can have a greater variety of configurations for the minimal area since the defect opens the possibility to have also cylindrical surfaces that start from the boundary of AdS and attach to the D5-brane. 
\vspace{0.2cm}\\ 
To investigate the possible configurations that can dominate
 for different values of $h/R$, we can start by fixing all the independent parameters in \eqref{parameters} except $h/R$, in such a way that the first loop is kept at a fixed distance from the defect while letting the second one to move away from it. In our parametrization the second circle is placed at a distance from the defect given by
\be
\frac{L}{R}+\frac{h}{R}\,.
\ee
\begin{figure}[t!]
	\centering
	\includegraphics[width=.95
	\textwidth]{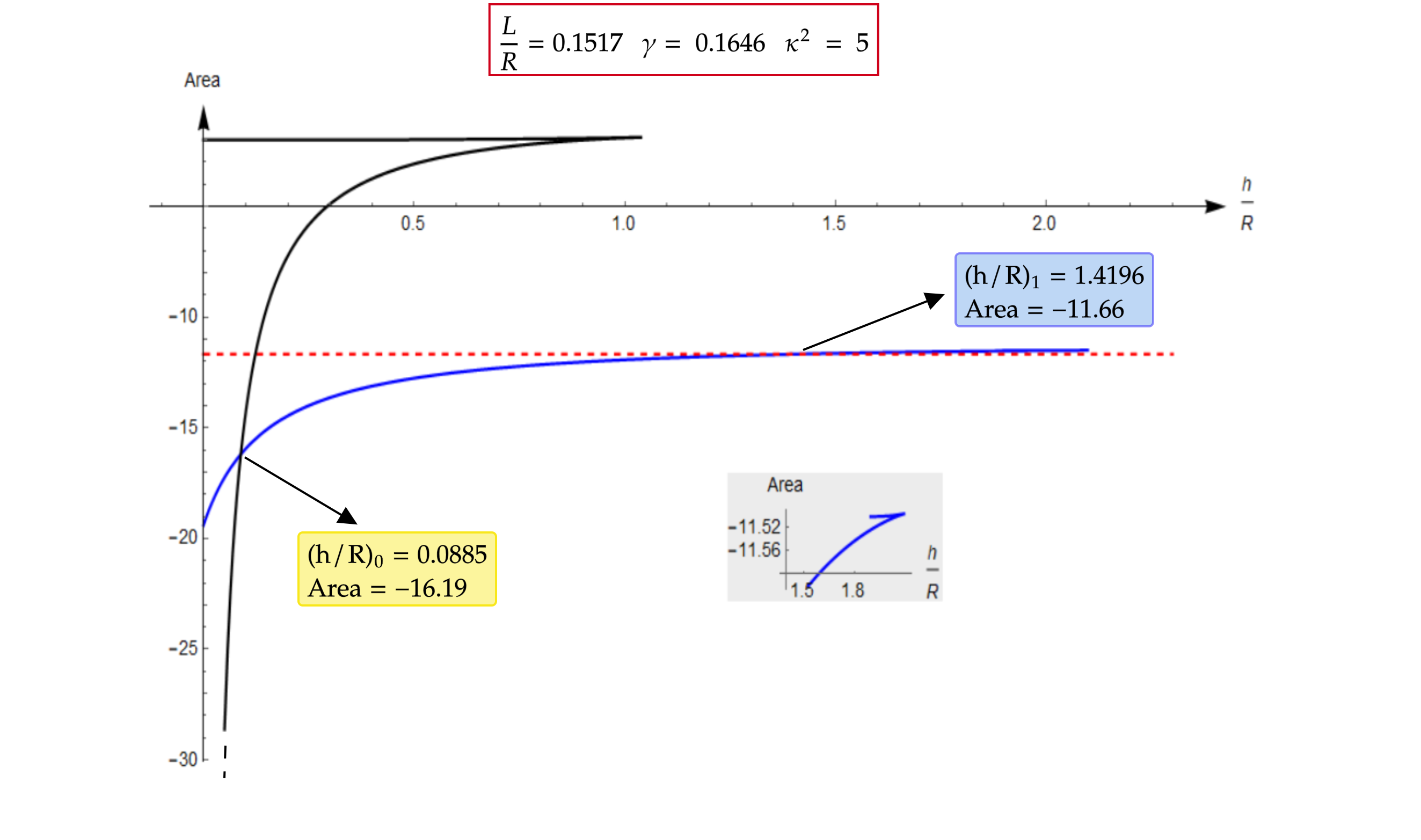}
	\caption{\footnotesize{ We have plotted the sum of the areas of the cylindrical surfaces for the two circles (blue curve) and the connected solution (black curve) as functions of the separation $h/R$ between the two loops at fixed $\k^2,\chi_1,\chi_2$ and $L/R$. The dashed red line corresponds to the configuration in which for the first loop we have a cylindrical surface attached to the D5-brane with a fixed area $\hat{S}_1= -10.66$ and the dome solution for the second loop (the area of the dome is normalized to -1), namely when $\hat{S}_1+\hat{S}_2=-11.66$. In the yellow box, we reported the value $(h/R)_0$ at which occurs the transition between the connected and the attached-attached configuration. In the light-blue box, we reported the value $(h/R)_1$ at which occurs the transition from the attached-attached configuration to the attached-dome one. }}
	\label{fig6}
\end{figure}
\noindent
In \autoref{fig6}, we plot the different types of solutions that are dominant for different values of $h/R$. We have chosen for the scalar couplings and the flux the values $\chi_1 = 0.7799,\,\chi_2=0.6153$ and $\kappa^2=5$. To show the behavior of the action as a function of $h/R$, we have also fixed the distance of the first loop from the defect $L/R=0.1517$. For this value, the cylindrical configuration extending from the boundary of AdS to the D5-brane is the solution with the minimal area for the first loop and, since we do not vary $L/R$, the value of $\hat{S}_1$ does not change. Approaching from infinity the defect with the second loop, we encounter a value of $h/R$ at which the cylindrical solution starts to exist also for the second circle, and at $(h/R)_1$ its area becomes equal to the area of the dome. Thus, for $h/R\leq(h/R)_1$ the minimal area configuration for the two loops is given by two cylindrical surfaces, one for each circle, attached to the defect. We refer to this solution as the attached-attached configuration. Keeping to decrease $h/R$, we reach the value  $(h/R)_0$ at which the area of the two cylindrical surfaces (blue curve in \autoref {fig6}), becomes equal to the area of the connected configuration between the circles (black curve). This configuration is visualized in \autoref{fig8yellowbox}. For smaller values of the separation distance, the latter becomes dominant with respect to the former. In the example reported in \autoref{fig6}, the connected solution becomes dominant for $h/R\leq 0.0885$. We can summarize as follows, in terms of the separation distance, the two different transitions that can be present in the two circles correlator when a defect is inserted in the theory
 \begin{itemize}
 	\item[\textbf{1)}]  $\mathbf{(h/R)_0:}$ transition between the connected and the attached-attached to the defect configuration
 		\item[\textbf{2)}] $\mathbf{(h/R)_1:}$ transition between the attached-attached and the attached-dome configuration.
 \end{itemize}
In the following, we will study different values of the physical parameters that will lead to different combinations of GO-like phase transitions.
\begin{figure}[t!]
	\centering
	\includegraphics[width=0.8
	\textwidth]{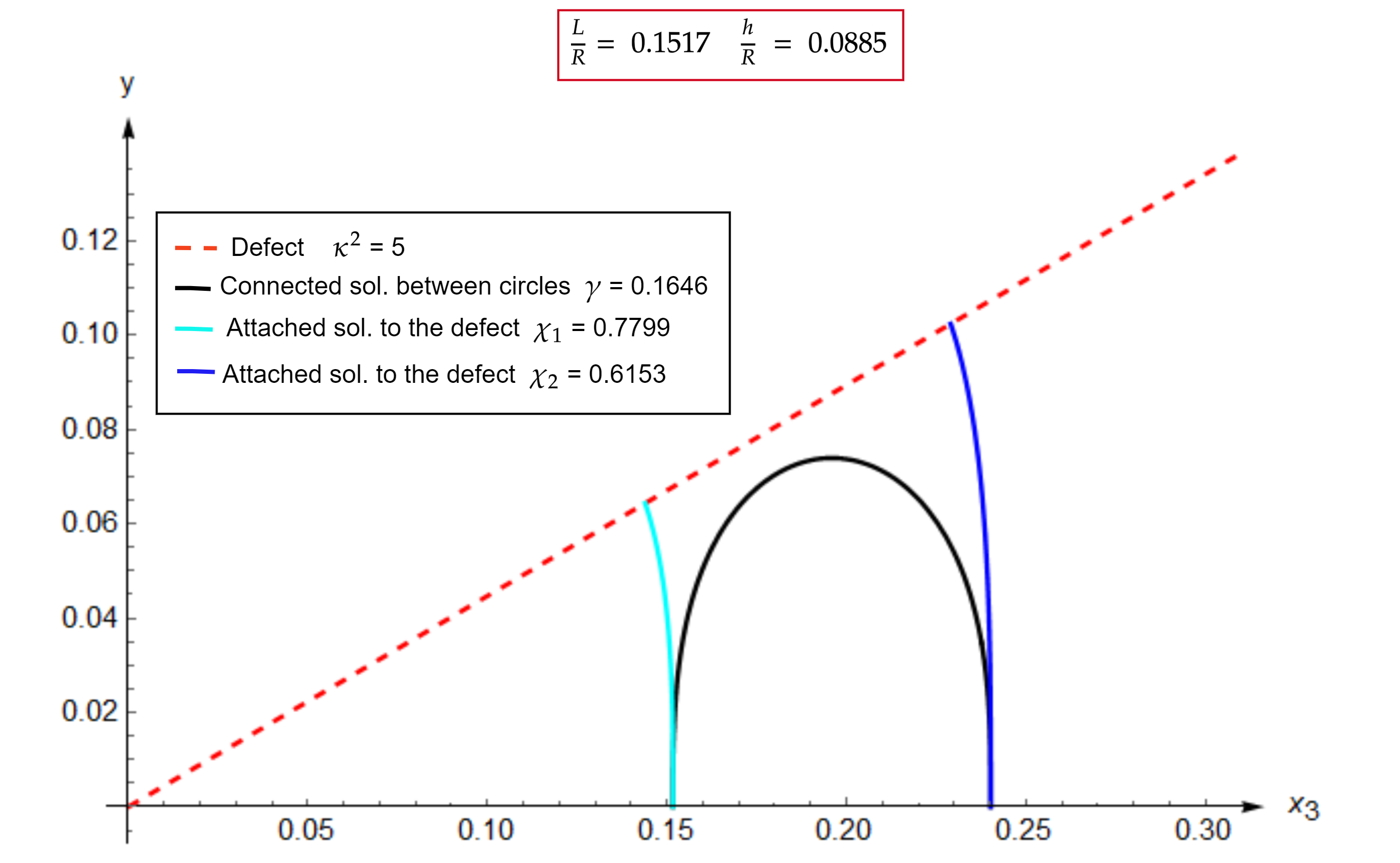}
	\caption{\label{fig8yellowbox}\footnotesize{We have plotted in the $x_3-y$ plane the defect (red line), the connected solution (black curve) and the attached-attached configuration (cyan and blue curves) at the transition point between the two different solutions corresponding to the yellow box in \autoref{fig6}.
	}	}
\end{figure}
\begin{figure}[h!]
	\centering
	\includegraphics[width=0.9
	\textwidth]{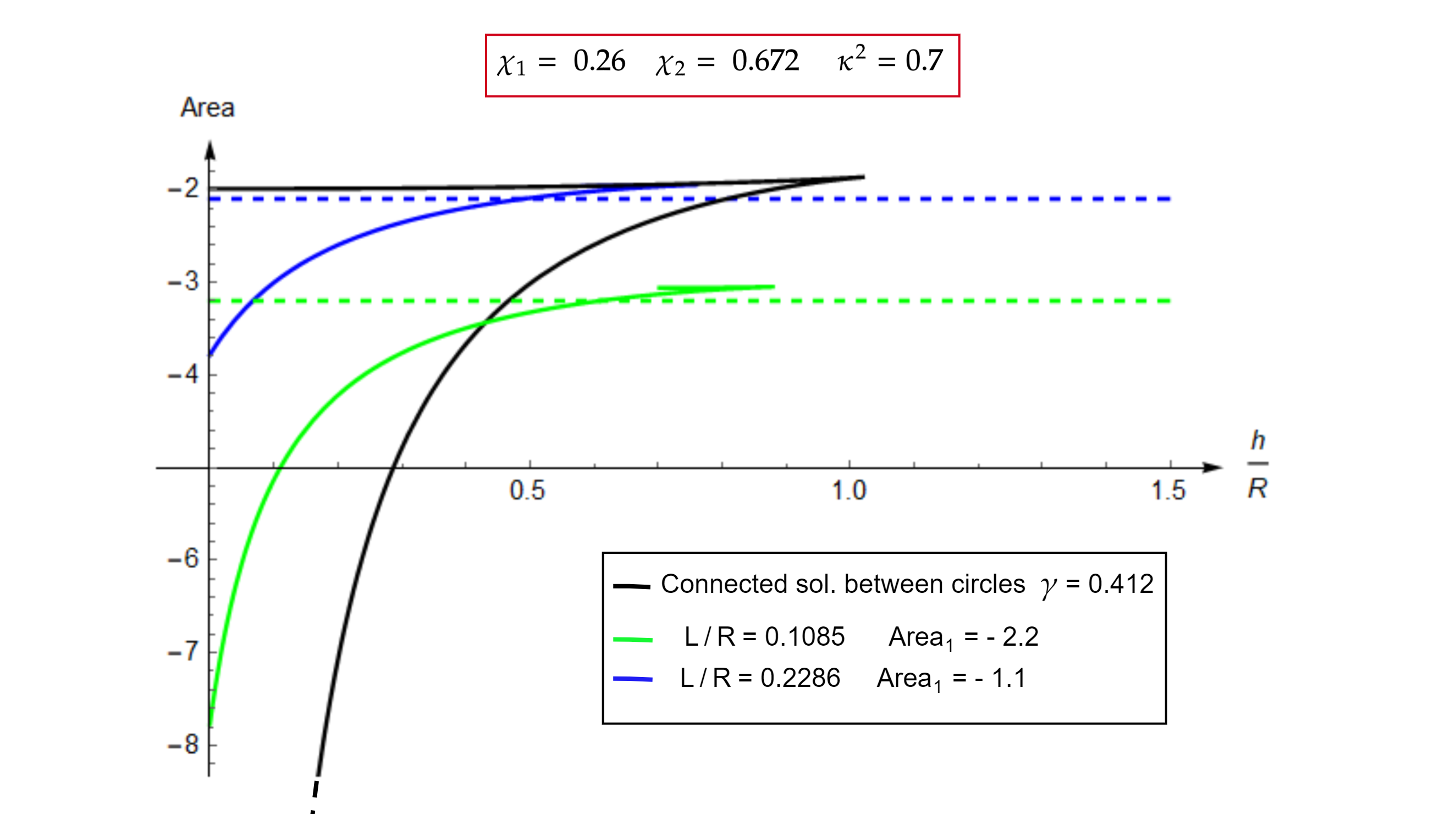}
	\caption{\footnotesize{We vary the distance from the defect of the first loop keeping fixed $\chi_1$, $\chi_2$ and $\kappa^2$. If the position of the first loop is $L/R=0.1085$, the connected solution (black curve) is dominant for $0 \leq h/R \leq 0.43$, then the minimal area configuration is given by two cylindrical surfaces attached to the defect (green curve). At $h/R=0.6096$ there is the second transition to the attached-dome configuration. The blue curve corresponds to put the first loop further away from the defect. In this case, there is only one relevant transition that is the one between the connected solution and the attached-dome configuration at $(h/R)_1=0.8069$.
}	}
	\label{fig7}
\end{figure}
\subsection{Different distances from the defect}
In \autoref{fig7}, we analyze what happens if we put the first circle at different positions with respect to the defect, without changing $\chi_1,\,\chi_2$, and $\kappa^2$. The green and blue curves correspond to the area of the attached-attached configuration for two different values of $L/R$. When the first loop is closer to the defect (green curve), we find two different types of Gross-Ooguri transitions as described before. For small values of $h/R$, the dominant solution is the connected one (black curve). Increasing the separation between the loops, we have a value of $h/R$ where the connected solution and the attached-attached configuration have the same area, corresponding to the black curve crossing the green one in \autoref{fig7} . If we increase further $h/R$, the attached-attached configuration becomes the minimal one, but the area of the cylindrical surface for the second circle becomes less negative since we are moving the loop away from the defect. When $\hat{S}_2=-1$, the second Gross-Ooguri transition to the attached-dome configuration takes place. That corresponds to have the cylindrical surface for the first circle, whose position is kept fixed, and the dome solution for the second one. 
In \autoref{fig7}, this transition occurs at the crossing point between the dashed-green line and the continuous green curve. When we increase $L/R$, the sum of the areas of the two cylindrical solutions becomes less negative. Thus, the
only relevant transition may be between the connected and the attached-dome configuration since there are no values of $h/R$ for which the attached-attached solution is dominant. This case is shown in \autoref{fig7} by the blue curve. The black curve does not change as we vary $L/R$, since it depends only on $\gamma$ and $h/R$.
\begin{figure}[t!]
	\centering
	\includegraphics[width=1.0
	\textwidth]{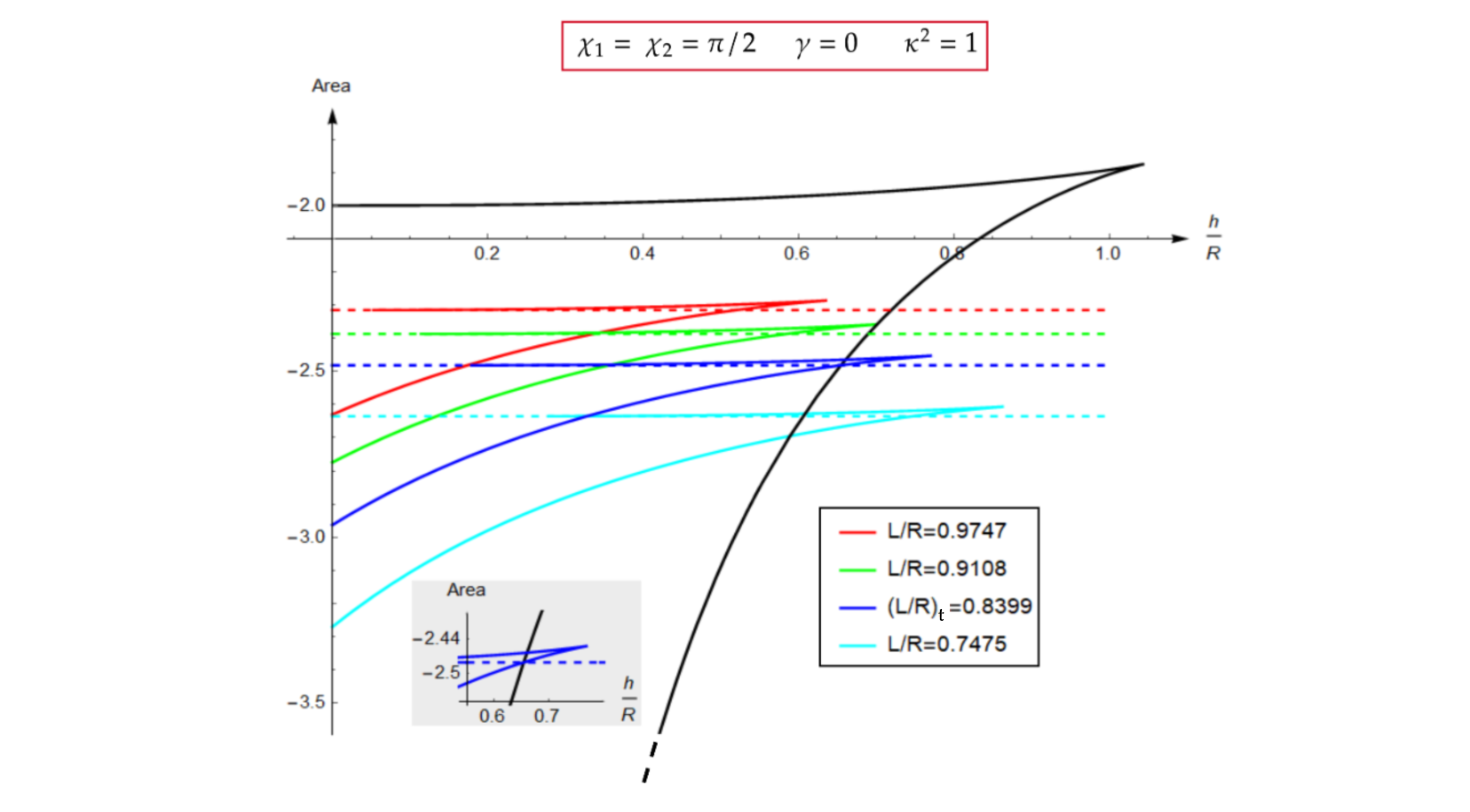}
	\caption{\footnotesize{We keep fixed $\k^2,\; \chi_1$ and $\chi_2$. The continuous and dashed colored curves correspond to the attached-attached and the attached-dome configurations, respectively. The black curve represents the connected solution between the circles. For $L/R=(L/R)_t$, there is a value of $h/R$ corresponding to the intersection between the continuous black and blue curves, and the dashed blue line at which the three different configurations have the same area. The blue curve separates situations in which two different types of Gross-Ooguri phase transitions are present ($L/R<(L/R)_t$), from cases in which the only relevant phase transition is between the connected and the attached-dome solution. In this plot, we have chosen $\chi_1=\chi_2=\pi/2$ since for these values of the angles is easier to determine numerically $(L/R)_t$. }}
	\label{triple}
\end{figure}
 There should be an intermediate case that interpolates between the two different behaviors shown in \autoref{fig7} and which is characterized by a certain value of $(L/R)$, denoted as  $(L/R)_t$, at which the connected, the attached-attached, and the attached-dome configurations have the same area for the same value of $h/R$.  We can summarize the different situations choosing different values for the distance from the defect, as follows
\begin{itemize}
	\item  $\boldsymbol{(L/R)<(L/R)_t}:$ there are two different types of transitions. At $(h/R)_0$ takes place the first one between the connected and the attached-attached configuration, while at $(h/R)_1$ the attached-dome solution becomes dominant (crossing between the cyan continuous curve and the dashed one in \autoref{triple}).
	\item  $\boldsymbol{(L/R)=(L/R)_t}:$ in this case $(h/R)_0=(h/R)_1$, namely the connected solution is dominant until the separation between the circles reaches a value at which this saddle-point has the same area as the attached-attached and the attached-dome solutions. We refer to this particular configuration as the triple point, since the three solutions have the same area for the same value of $h/R$. Keeping to increase $h/R$, the solution with the minimal area is given by a cylindrical surface attached to the defect for the loop located at $(L/R)_t $, and the dome for the second one. 
	\item  $\boldsymbol{(L/R)>(L/R)_t}:$ the attached-attached configuration is never dominant and the only relevant transition is the one between the connected and the attached-dome solutions. 
\end{itemize} 
\subsection{Transitions for different fluxes and angles}
\begin{figure}[t!]
	\centering
	\includegraphics[width=0.9
	\textwidth]{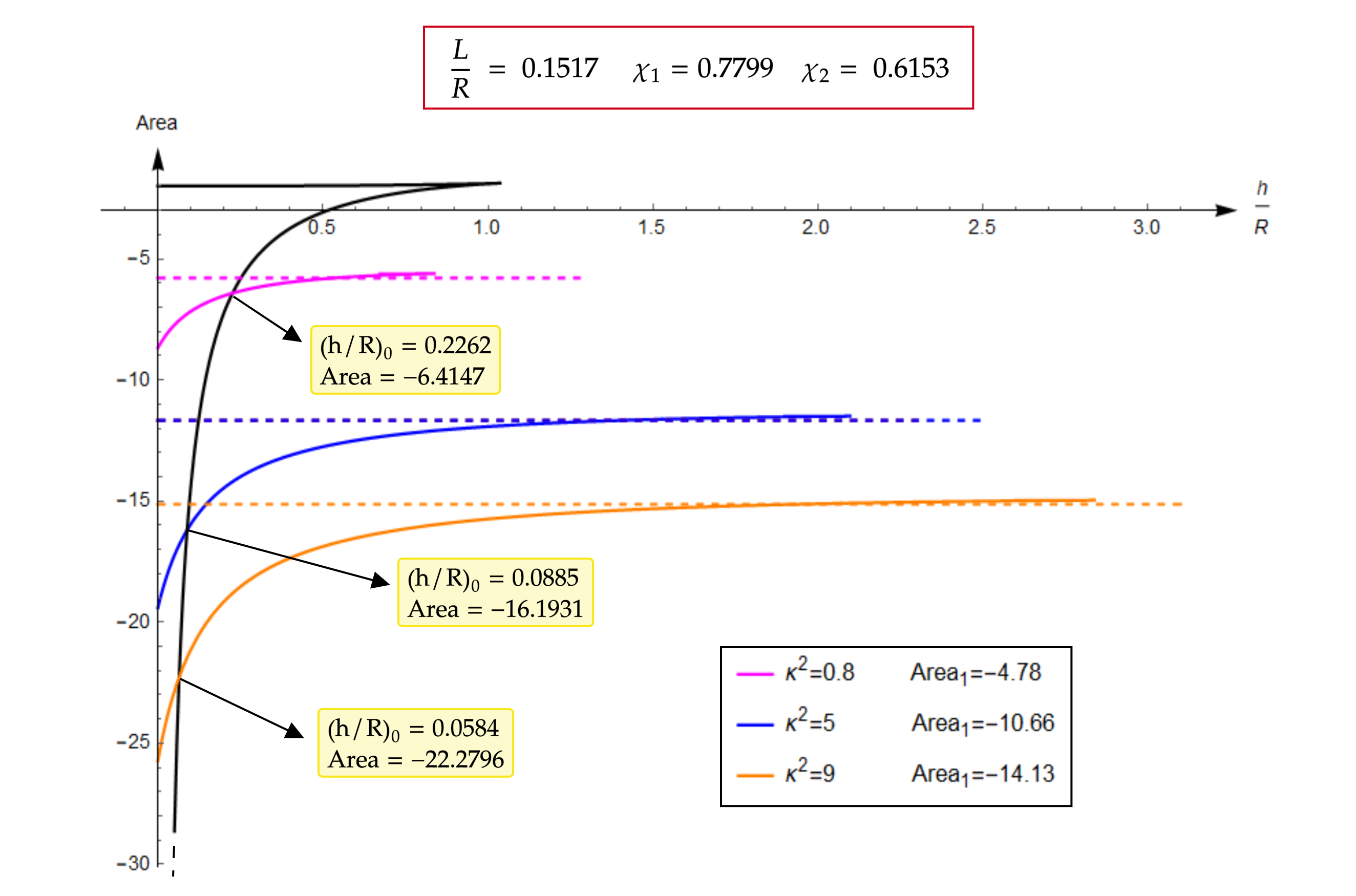}
	\caption{\footnotesize{The distance from the defect of the first loop and the angular separation $\gamma=0.1646$ are kept fixed. The connected solution between the circles is depicted by the black curve. At $(h/R)_0$ occurs the transition between the connected configuration and the attached-attached solution, since the minimal area of the first loop with $\chi_1=0.7799$ at $L/R=0.1517$ admits a cylindrical surface attached to the defect for the values of the flux considered.} Increasing $\kappa^2$, the value of $(h/R)_0$ decreases while $(h/R)_1$, the distance at which occurs the transition from the attached-attached to the attached-dome configuration, increases.}
	\label{fig8}
\end{figure}
\noindent
In \autoref{fig8}, we plot different types of solution for fixed $\chi_1,\,\chi_2$ and $L/R$, varying the flux. 
The connected configuration does not change since it is independent on $\kappa^2$. For the values of $\kappa^2$ and $L/R$ considered, the cylindrical solution for the first circle is dominant with respect to the dome.
 The transition between the connected and the attached-attached configuration is possible and it occurs at a value of the separation distance between the loops, given by $(h/R)_0$, that decreases as the flux grows since $\hat{S}_1+\hat{S}_2$ becomes more negative. On the other hand, the space of parameters of the cylindrical solution is enlarged by increasing the flux (see fig. 4 of \cite{Bonansea:2019rxh}). Thus, the value $(h/R)_1$ where occurs the transition to the attached-dome configuration, increases with the flux. When $\k^2$ grows, the attached-attached solution dominates for a larger range of values of $h/R$, because the slope of the brane becomes smaller and the defect gets closer to the boundary. As we increase the flux, it is energetically preferred for the system to settle in the attached-attached solution.  \vspace{0.3cm}\\ 
\begin{figure}[t!]
	\centering
	\includegraphics[width=1.0
	\textwidth]{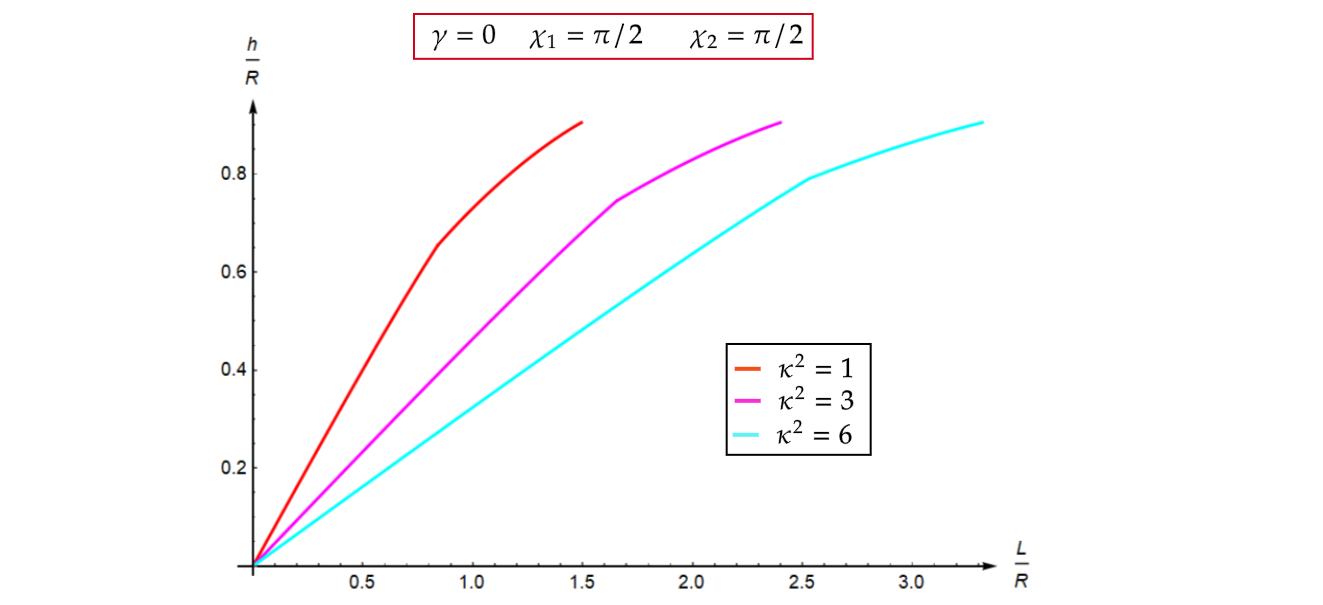}
	\caption{\footnotesize{ For $\chi_1=\chi_2=\pi/2$, we plot how the values of $h/R$, at which the first relevant transition takes place, change as we vary both $\kappa^2$ and $L/R$. The variation in the slope of the curves corresponds to the triple point at which a change in the type of the first relevant transition takes place. For small $h/R$, the first transition that we encounter is the one between the connected and the attached-attached configuration. If the first loop is placed at $L/R>(L/R)_t$, the transition between the connected and the attached-dome solution becomes the first relevant one. The curves stop for the value of $L/R$ at which the cylindrical surface for the first loop becomes less dominant with respect to the dome. For larger distances from the defect of the first loop, the only relevant transition is standard GO, that for $\gamma$=0 occurs at $h/R=0.91$. Increasing the flux, the curves terminate at larger $L/R$ because the cylindrical surface is dominant with respect to the dome for a greater interval in $L/R$.}}
	\label{transition}
\end{figure}
\noindent
In \autoref{transition}, for fixed $\chi_1\;,\chi_2,$ varying $L/R$ and $\kappa^2$, we plot the separation distance between the loops at which the first relevant transition occurs. For small values of $h/R$, the first relevant transition is the one between the connected and the attached-attached configuration. A change in the slope of the curves takes place at $L/R=(L/R)_t$, namely when the triple point is reached. For $L/R>(L/R)_t$, the attache-attached solution is never dominant and the first relevant transition is between the connected and the attached-dome configuration. In this plot, it is better displayed that, as the flux is increased, the transition between the connected and the attached-attached solution occurs at smaller $h/R$ for equal values of $L/R$. Moreover, we can also notice that $(L/R)_t$ increases with the flux. All the curves stop when the loop closer to the defect is placed at a value of $L/R$ such that the cylindrical surface attached to the defect becomes less preferred with respect to the dome. Thus, for larger $L/R$, the only relevant transition is between the connected surface and the two domes, as in the standard GO case. For $\gamma=0$, it happens at $h/R=0.91$, which is the value at which all the curves stop in \autoref{transition}. These considerations are displayed in \autoref{phasediagram}, where we plot for fixed $\kappa^2$, $\chi_1$ and $\chi_2$ the scheme of the possible transitions in the plane of the parameters $(L/R,h/R) $. 
\begin{figure}[t!]
	\centering
	\includegraphics[width=0.96
	\textwidth]{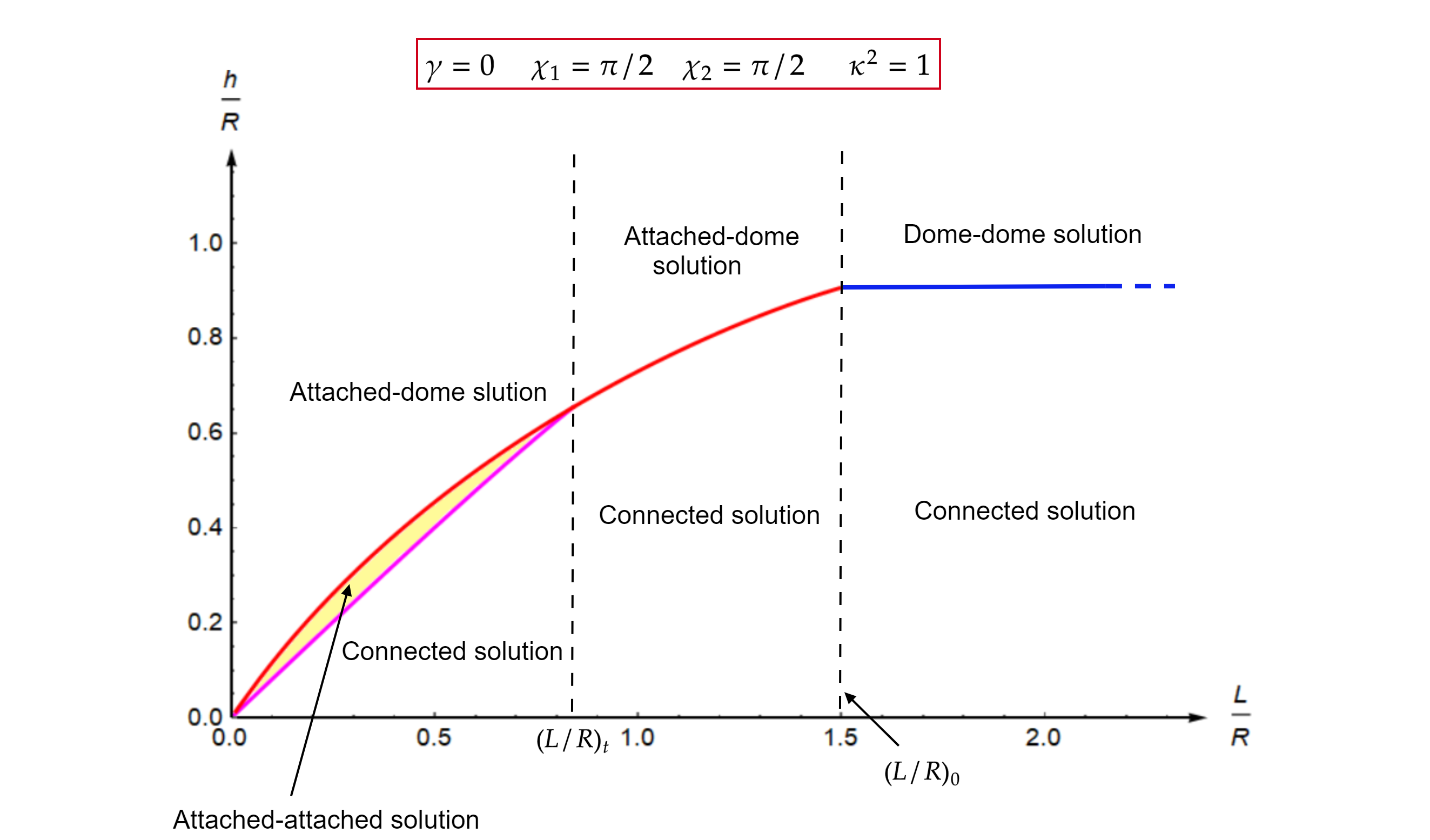}
	\caption{\footnotesize{For fixed $\kappa^2$, $\chi_1$ and $\chi_2$ we plot in the $(L/R,h/R)-$plane the possible phase transitions that can occur depending on the region of parameters that we are considering. When $(L/R) <(L/R)_t$, the magenta line represents the values of $h/R$ at which the transition between the connected and the attached-attached configuration takes place. The red line corresponds to the transition between the attached-attached and the attached-dome solutions. The connected configuration is dominant below the magenta line while the attached-attached solution has the minimal area in the yellow region. Above the red line, the attached-dome configuration becomes the dominant one.  For $(L/R)_t< (L/R) <(L/R)_0 $, there is only one relevant transition depicted by the red line. The value $(L/R)_0 $ corresponds to the distance at which the cylindrical surface attached to the defect for the first circle becomes less preferred with respect to the dome. Thus, for $(L/R) \geq (L/R)_0 $, the only transition is the standard GO one between the connected solution and the two separated dome surfaces, which occurs at the constant value $h/R=0.91$ indicated by the blue line.}}
	\label{phasediagram}
\end{figure}
\vspace{0.3cm}\\ 
Now, we want to analyze what happens if we exchange $\chi_1$ and $\chi_2$. Contrary to the case without the defect, where only the difference between the two is relevant, here also the value of the single angle is important. This is because the defect breaks the original $SO(6)$ R-symmetry of $\mathcal{N}=4$ SYM theory to $SO(3) \times SO(3)$. In \autoref{fig9}, we have fixed two values for the coupling with the scalars, $\kappa^2$, and $L/R$. The connected solution (black curve) depends only on $\gamma$ and it is invariant under the exchange of $\chi_1$ and $\chi_2$. 
\begin{figure}[t!]
	\centering
	\includegraphics[width=0.9
	\textwidth]{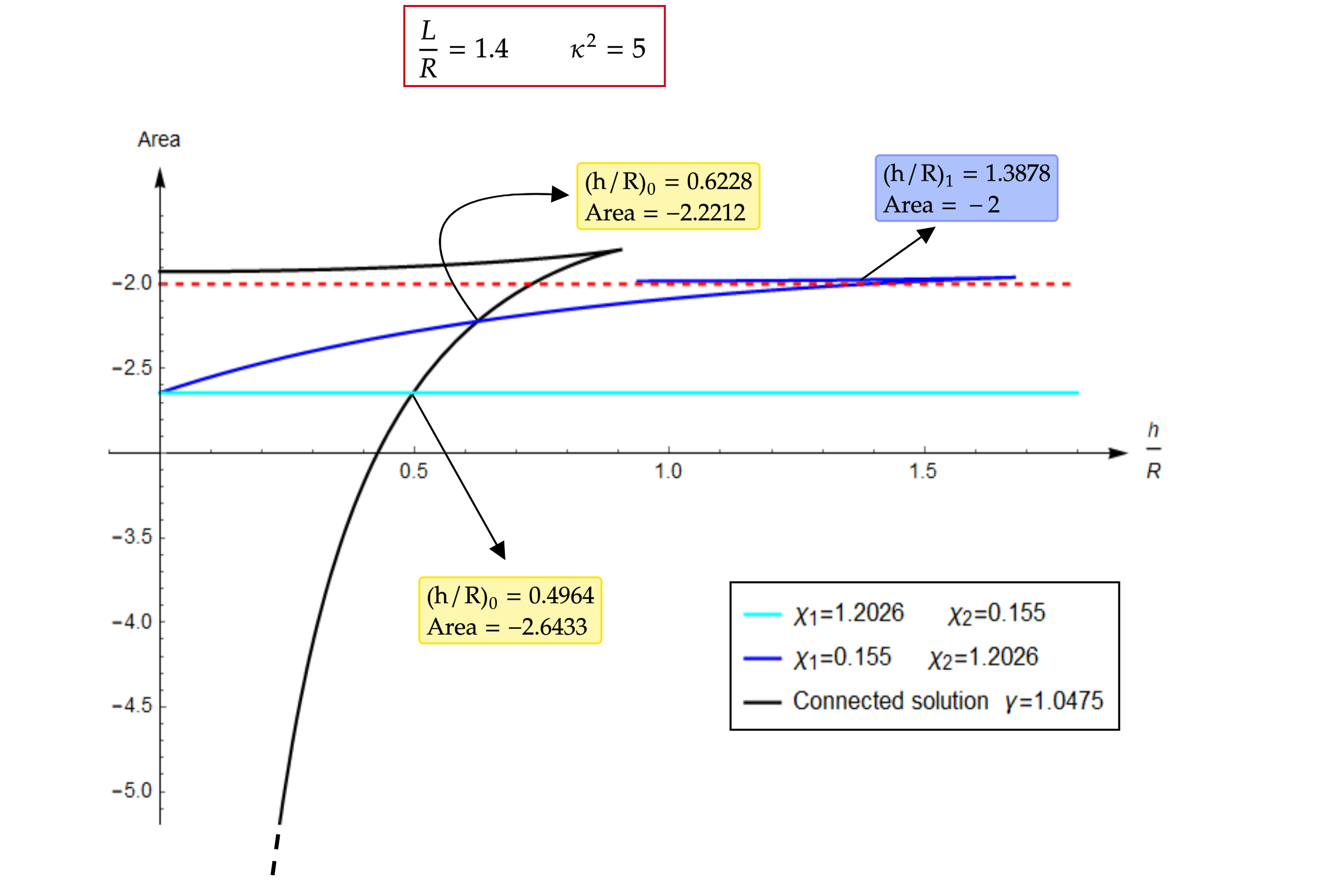}
	\caption{\footnotesize {
			We have fixed $L/R=1.4$. For the circle with the smallest angle, this value of $L/R$ is such that no cylindrical surface attached to the defect exists. Thus, if  $\chi_1=0.155$ and $\chi_2=1.2026$, we have that for $h/R \leq 0.6288$ the dominant solution is the connected one between the two circles (black curve). Then, the dome-attached configuration will start to be dominant (blue curve), until $h/R$ reaches $(h/R)_1 =1.3878$. The dashed red line corresponds to the area of two domes, namely $\hat{S}_{\text{ren}}=-2.$ When we exchange the values of $\chi_1$ and $\chi_2$, the loop closer to the defect has the largest angle. The cyan line depicts the attached-dome configuration in this case where the connected solution has the minimal area until $(h/R)_0 \leq0.4964$. For larger values of $h/R$, the attached-dome confiuguration becomes dominant and this is the only transition that occurs if we keep fix the distance from the defect of the first loop. }}
	\label{fig9}
\end{figure}
We can choose a value for $L/R$ such that no cylindrical surface exists for the loop with the smaller angle, while for the second circle the cylindrical surface exists and its area is smaller than the area of the dome, as is shown in \autoref{fig10}.
If $\boldsymbol{\chi_1<\chi_2}$, we can have the following situations depending on the value of $h/R$:
\begin{itemize}
	\item[$\boldsymbol{(1)}$] $\boldsymbol{0\leq h/R\leq(h/R)_0:}$ the connected solution is the dominant one.
	\item[$\boldsymbol{(2)}$] $\boldsymbol{(h/R)_0< h/R\leq(h/R)_1:}$ the configuration with the minimal area is the dome-attached solution, represented in \autoref{fig9} by the blue curve. The loop which is closer to the defect is characterized by the angle $\chi_1$. Thus, only the dome solution is possible at the chosen value of $L/R$, while for the circle with $\chi_2$ the cylindrical surface is dominant with respect to the dome.
	\item[$\boldsymbol{(3)}$] $\boldsymbol{(h/R)>(h/R)_1}:$ at $(h/R)_1$ the area of the cylindrical surface for the second circle becomes equal to -1 and the dome-dome configuration, represented in \autoref{fig9} by the dashed-red line, is the one with the minimal area. Compared to the case without the defect, the transition to the dome-dome configuration occurs at a larger value of the separation distance.
\end{itemize}
If we exchange the values of the angles in such a way that $\boldsymbol{\chi_1>\chi_2}$, we have only two possible situations:
\begin{itemize}
	\item[$\boldsymbol{(1)}$] $\boldsymbol{0\leq h/R\leq(h/R)_0:}$ the connected solution is the dominant one.
	\item[$\boldsymbol{(2)}$] $\boldsymbol{h/R>(h/R)_0:}$ the configuration with the minimal area is the attached-dome solution, given in \autoref{fig9} by the cyan line. This configuration has a constant area since we are moving away from the defect the loop whose solution, for this range of parameters, is the dome. Thus, its area remains constant and equal to -1 as we vary $h/R$.
\end{itemize}
In \autoref{fig9}, we can notice that exchanging $\chi_1$ and $\chi_2$ also the value of $h/R$ at which the connected solution stops to be dominant changes. Moreover, we can choose different pairs of $\chi_1$ and $\chi_2$ that give the same $\gamma$, but we do not expect the same result for the values of $(h/R)_0$ and $(h/R)_1$, as shown in \autoref{fig11}. When $\chi_1$ and $\chi_2$ are larger, the attached-attached configuration can exist and dominate for larger values of $h/R$.
\vspace{1cm}
\begin{figure}[h!]
	\centering
	\includegraphics[width=0.9
	\textwidth]{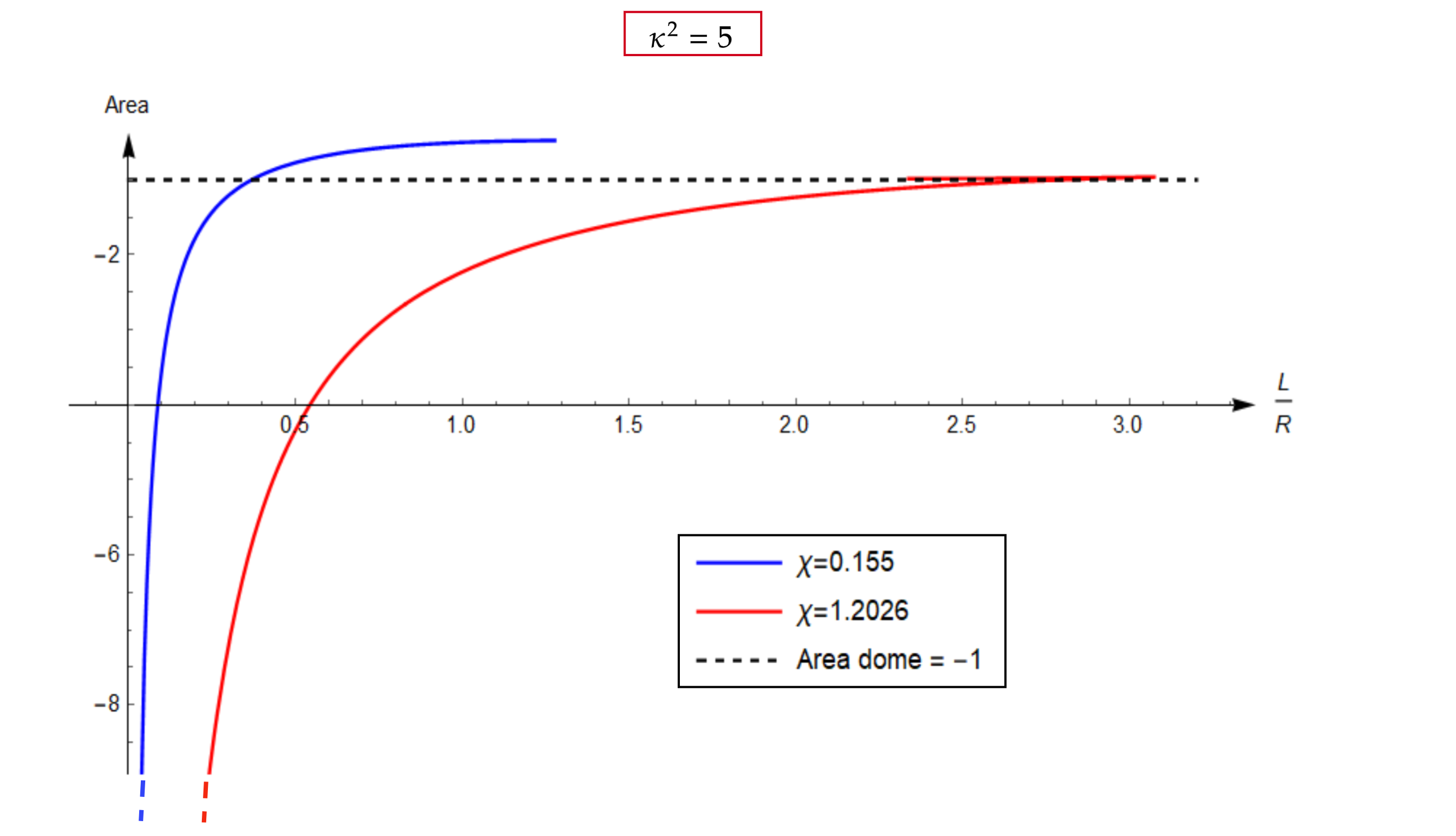}
	\caption{\footnotesize{ We plot the area of the cylindrical solution as a function of the distance from the defect for two different angles and for $\kappa^2=5$. In  \autoref{fig9}, we have chosen the position of the circle closer to the defect to be $L/R=1.4$. Thus, no cylindrical solution exist for the loop with $\chi=0.155$ (blue curve). For the other one (red curve), with $\chi=1.2026$, the area of the cylindrical solution at  $L/R=1.4$ is -1.6433 and dominant with respect to the dome configuration. }}
	\label{fig10}
\end{figure}	
\newpage
\begin{figure}[t!]
	\centering
	\includegraphics[width=0.9
	\textwidth]{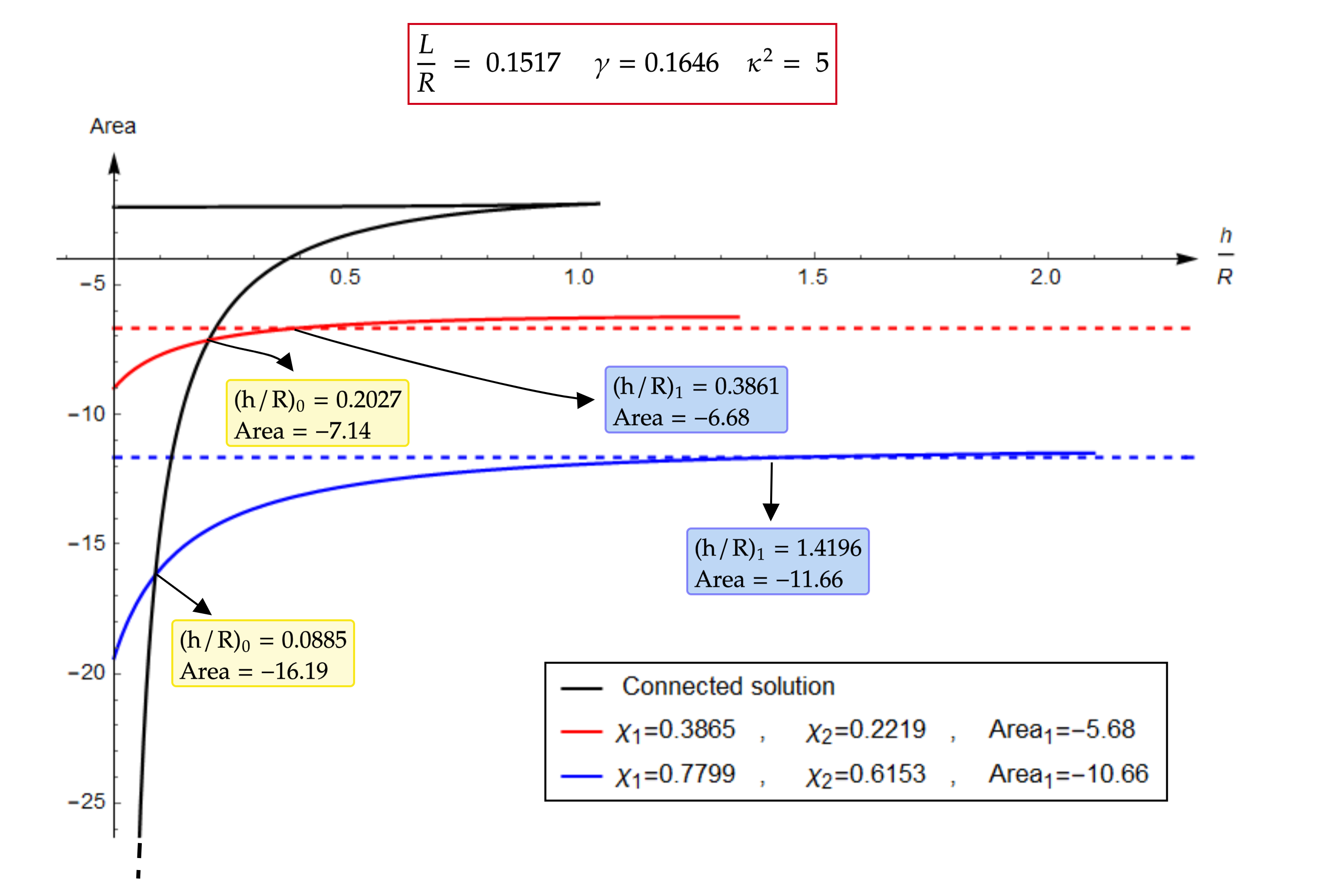}
	\caption{\footnotesize{Fixing $L/R$ and $\gamma$, we consider two different pair of $\chi_1$ and $\chi_2$ such that they correspond to the same $\g$. If we take larger values for the two angles (blue curve), the transition between the connected solution and the attached-attached configuration occurs at a smaller $(h/R)_0$, compared to the case depicted by the red curve. The second transition between the attached-attached and attached-dome solution occurs at a greater $(h/R)_1$ if $\chi_1$ and $\chi_2$ are larger. }}
	\label{fig11}
\end{figure}
\subsection{String-brane crossing}
\begin{figure}[t!]
	\centering 
	\includegraphics[width=.90\textwidth]{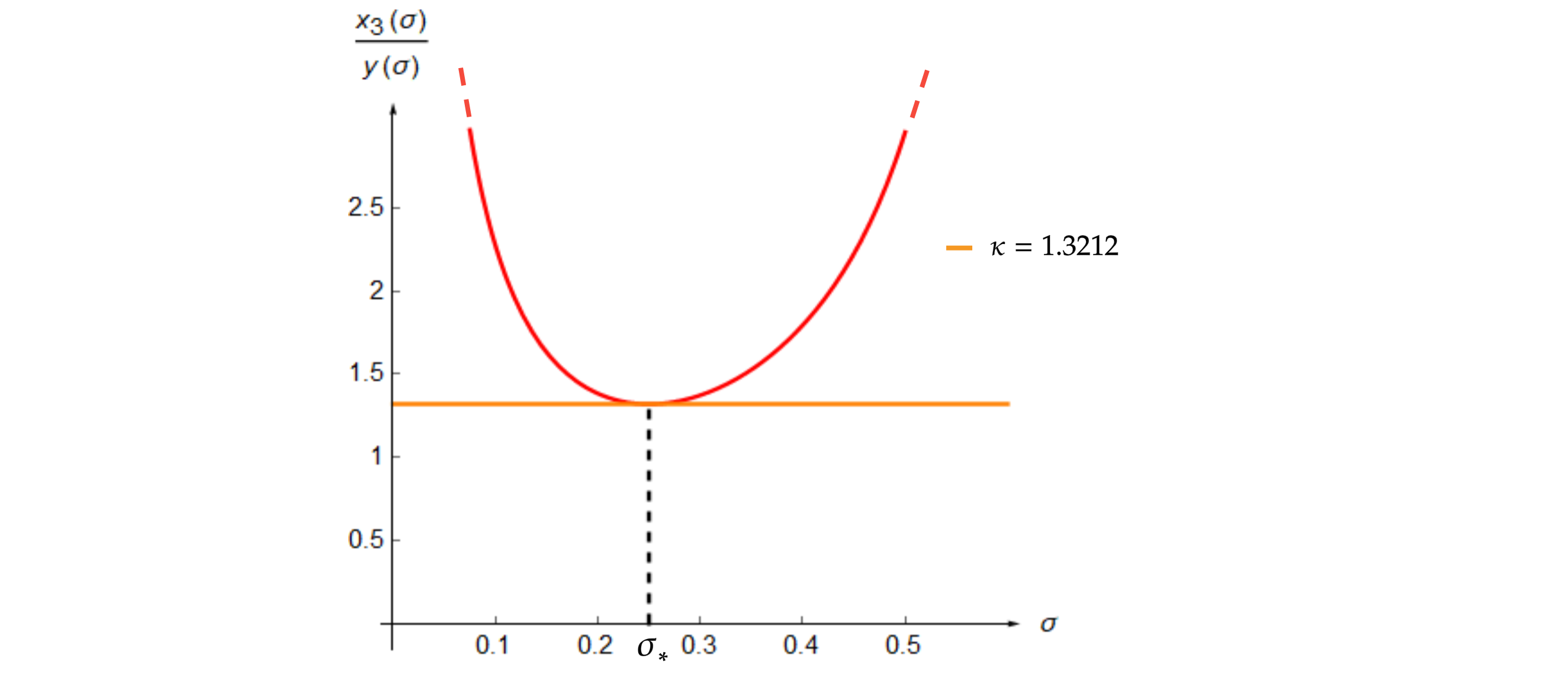}	
	\caption{\label{fig17Ushaped}\footnotesize{The equation $x_3(\s)/y(\s)=\k$ has a unique solution when $\s=\s_*$, which corresponds to the minimum of $x_3(\s)/y(\s)$ (red curve) and to the connected solution between the circles being tangent to the defect. In the figure, we plot  $x_3(\s)/y(\s)$ selecting $L/R=0.2184\;,\chi_1=0.8$ and $\chi_2=2.8$. The value of the flux for which the brane is tangent to the defect is $\k_*=1.3212$ (orange line).}}
\end{figure}

In this section, we will inspect whether or not the connected solution between the loops can be intersected by the defect. If this is feasible, the transition between the connected and the attached-attached configuration can happen before the latter becomes dominant with respect to the former and a zero-order phase transition can occur. \\
 The D5-brane has the following profile inside $AdS_5\times S^5$ \cite{Nagasaki:2011ue,Nagasaki:2012re}
\be
\label{defsol}
y=\frac{x_3}{\kappa},\quad\quad \theta=\frac{\p}{2},\quad\quad \alpha_2=\alpha^{(0)}_2\quad\quad\mathrm{and}
\quad\quad\beta_2=\beta^{(0)}_2,
\ee
where $\alpha_2^{(0)}$ and $\beta_2^{(0)}$ are two constant values. If the defect is tangent to the connected solution in $\s_*$, the following conditions have to be satisfied
\begin{align}
\label{cond1}
AdS_5\; &\text{part:}\quad \frac{x_3(\s_*)}{y(\s_*)}=\k,
 \\ 
\label{cond2}
S^5\; &\text{part:}\quad \theta(\s_*)=j\s_*+\chi_1=\frac{\pi}{2}\,.
\end{align}
Eq. \eqref{cond1} has a unique solution in the interval $\s_*\in [0,\;\hat{\s}/2 ] $ when $\s_*$ corresponds to the minimum of the function $x_3/y$.
 Thus, we have a further condition to impose 
\be
\label{condder}
\frac{d}{d\s}\left( \frac{x_3(\s)}{y(\s)}\right)\bigg|_{\s=\s_*}=0\,,
\ee
which can be seen, for fixed values of $L/R$ and the angles, in \autoref{fig17Ushaped}.
\noindent 
This derivative is always positive in the interval $(\hat{\s}/2,\;\hat{\s})$, where it does not possess a minimum.
Solving eq. \eqref{cond2}, we can express $\s_*$ as a function of $\chi_1,\;\chi_2$ and $m$ as
\be
\s_*=\left(\frac{\pi}{2}-\chi_1 \right) \frac{\sqrt{\g^2-4\mathds{K}(m)^2(m+1)}}{\g}\;.
\ee
Since $\s_*$ has to be positive, we assume $\chi_1<\pi/2$. The condition $\s_* \leq \hat{\s}/2$ gives an additional constraint on the values that $\chi_2$ can assume, namely  $\pi-\chi_1 \leq\chi_2 \leq \pi$.
Combining eqs. \eqref{condder} and \eqref{cond1}, we can determine the critical value of the flux at which the connected solution touches the brane
\be
\kappa_*=\frac{\sqrt{\e_0}}{R}\frac{y^2(\s_*)}{y'(\s_*)}\sech \left( \frac{v(\hat{\s})}{2}\right)\;,
\ee
where we have used the equation of motion for $x_3$ 
\be
x'_3(\s)=cy^2(\s) \qquad \text{with}\quad c=\frac{\sqrt{\e_0}} {R}\sech\left(\frac{v(\hat{\s})}{2} \right)\,.
\ee
Due to $\s_*\leq\hat{\s}/2$, $y'$ is positive \footnote{With $y'$ we indicate the derivative of $y(\s)$ respect to $\s$.} and $\k_*>0$. Using eq. \eqref{cond1}, we get an expression for the critical distance $\left( \frac{L}{R}\right) _*$ of the first circle from the defect at which the worldsheet stretching between the two circles and the brane touch each other. \newpage 
The explicit form of $\left( \frac{L}{R}\right) _*$ and $\kappa_*$ is displayed in appendix \ref{appB}.
The critical parameters $\left( \s_*,\;L_*/R_*,\;\k_*\right) $ can be expressed as functions of $\chi_1,\;\chi_2$, and $m$ (that corresponds to a specific value for $h/R$). Thus, as we vary $h/R$, the values $\left( \s_*,\;L_*/R_*,\;\k_*\right) $ change.
\begin{figure}[t!]
	\centering 
	\includegraphics[width=.80\textwidth]{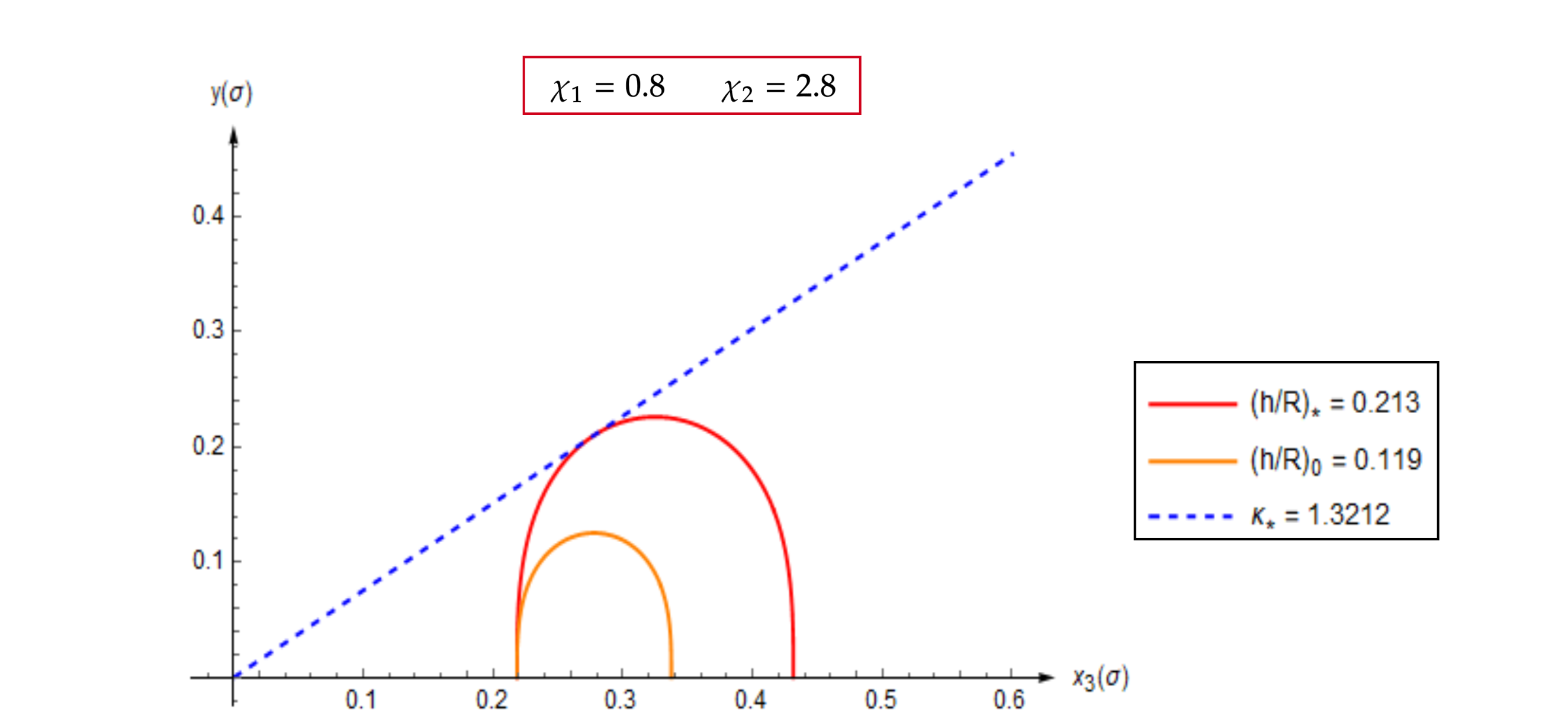}	
	\caption{\footnotesize{Fixed $L/R=0.2184$ and the two angles $\chi_1$ and $\chi_2$, we found the values of $\kappa_*$ and $(h/R)_*$ at which the connected solution touches the brane (which profile in the $x_3-y$ plane is given by the dashed blu line). The red curve depicts the connected solution touching the defect for $(h/R)_*=0.213$. The orange curve represents the value of $h/R$ at which the connected solution ceases to be dominant and, for these values of the parameters, the attached-attached becomes the one with smaller area.}}
	\label{touching}
\end{figure} 
\noindent
Alternatively, we can fix the two angles, the distance from the defect of the first circle with scalar coupling $\chi_1$, and numerically determine the set of values $(\k_*,\;h_*/R_*,\;\s_*)$ at which occurs the crossing. For $\k<\k_*$ or $ h/R<(h/R)_*$, keeping fixed the other parameters, the connected configuration remains below the brane without crossing. Increasing the distance from the defect of the connected solution, at fixed angles on the $S^5$, $\k_*$ and $(h/R)_*$ grow. Notice that to have a brane-string intersection in the $S^5$ part of the space, it is necessary that the angles $\chi_1$ and $\chi_2$ belong to different hemispheres, namely $\chi_1 \in (0,\pi/2]$ and $\chi_2 \in [\pi/2,\pi)$. The touching may take place at a value $(h/R)_*$ such that the connected solution is no longer dominant. In \autoref{touching} we show that, for the values of the parameters considered, the transition between the connected solution and the attached-attached configuration takes place for a separation distance between the circles smaller than $(h/R)_*$. Proving in general that the touching always happen when the connected configuration has no longer the minimal area is a hard task, due to the large parameter space involved. Thus, to be sure to avoid the string-brane crossing, we take $\chi_1$ and $\chi_2$ in the same hemisphere.  \vspace{0.2cm} \\
Moreover, we can focus the $\gamma=0$ case that corresponds to have no motion of the string in the $S^5$. It makes sense to analyze the problem of the string-brane touching for this value of $\gamma$ only when the angles are both equal to $\pi/2$. Otherwise, the brane and the defect are not tangent in the $S^5$ part of the space. The condition in eq. \eqref{cond2} is always satisfied if $\chi_1 =\chi_2 =\pi/2$,  since $j=0$. Therefore, we have to impose only the two constraints related to the touching in the $AdS_5$ part of the space, given in eq. \eqref{cond1} and \eqref{condder}. Through these conditions, we can fix the value of $(h/R)_*$ and $\s_*$ for which the connected solution and the defect become tangent. Keeping fixed $\k,\;\chi_1$ and $\chi_2$, we can find different values of $L/R$ and $h/R$ at which the connected solution touches the probe brane. Conversely, in the $\gamma \neq0$ case at a certain value of $\k_*$ corresponds only a possible value for $(L/R)_*$ and $(h/R)_*$ at which the touching occurs. In fact, there are other values of these parameters that satisfy the conditions in eq. \eqref{cond1} and \eqref{condder} in principle, but not the one in eq. \eqref{cond2} regarding the $S^5$ sphere. In \autoref{fig17}, varying the distance from the defect of the first loop and for fixed $\k$, and $\chi_1=\chi_2=\pi/2$, we plot the values of $h/R$ at which the transition between the connected and the attached-attached configuration occurs (blue line). The red curve displayes the values of $h/R$ corresponding to the connected solution being tangent to the defect. In the case considered, which is also investigated in \autoref{triple}, the transition always happen before the touching. We have numerically verified that this also happens for other values of the flux.   
\begin{figure}[t!]
	\centering 
	\includegraphics[width=.97\textwidth]{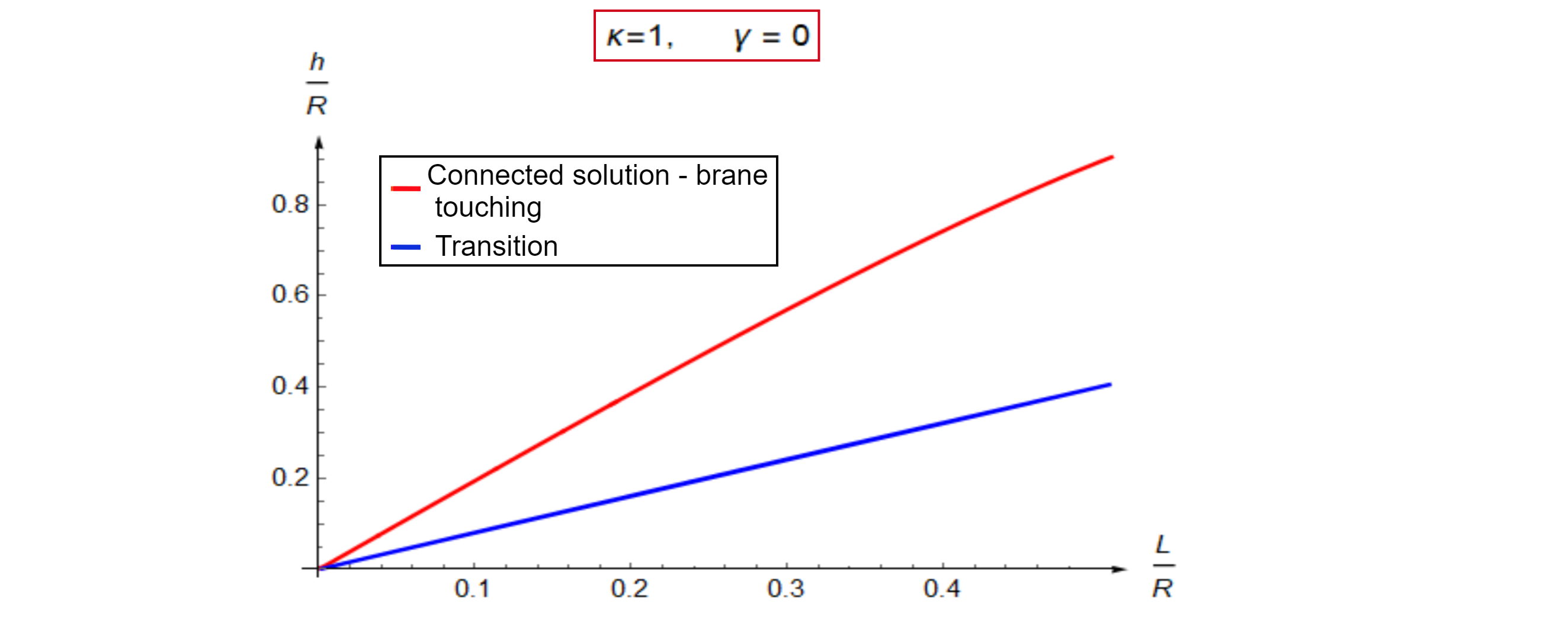}	
	\caption{\footnotesize{Fixed $\k=1$ and the two angles $\chi_1=\chi_2=\pi/2$, the red curve shows the values of $h/R$, varying the distance of the first loop from the defect, at which the connected solution touches the brane. The blue curve displays the values of the separation distances between the circles at which the connected solution ceases to be the minimal area configuration and the attached-attached one becomes dominant. For these values of the parameters, the transition always happens before the touching. If $L/R=0.507$, the touching occurs at $h/R=0.91$, namely when the connected solution experiences the transition to the two dome configuration. Thus, we do not consider larger values of $L/R$ in the plot because the touching would certainly happen when the connected solution is not the dominant one. }}
	\label{fig17}
\end{figure} 
\section{Conclusions and Outlook}
We have further made a study of the string theory dual of the two concentric circular Wilson loops correlator in a defect version of $\mathcal{N}=4$ SYM, in the equal radii case. The defect is realized as a probe D5 brane in the gravity side of the AdS/CFT correspondence. We analyzed how the GO phase transition \cite{Gross:1998gk}, that takes place in the correlator of two circular Wilson loops, is modified when a defect is inserted. 
When two circles are considered, the parameter space characterizing the different saddle-points that contribute to the evaluation of the correlator, is even richer than the single circle case. It is given by $L/R$, $h/R$, $\chi_1$ and $\chi_2$, and $\kappa$, namely the distance of the first loop from the defect, the separation distance between the loops, their angular positions in $S^5$, and the flux, respectively. Thus, the Wilson loops correlator can undergo new types of GO-like phase transitions whose analysis becomes more intricate. The standard GO-transition takes place when the separation distance between the two circles gets larger and is above a critical value at which the connected worldsheet solution between the two contours splits into two disjointed domes that have a smaller area, and that continue to interact through the exchange of light supergravity modes. In our case, this picture remains valid when we are far away from the defect, but when the circles get closer to it new minimal surface solutions are present. In particular, for some values of the parameters, the configuration with the minimal area can consist of two cylindrical surfaces attached to the defect or of one cylindrical surface for one of the contours, and a dome for the second one. 
We have studied the possible transitions between different minimal surface configurations in different regions of the parameter space. The solution corresponding to the string worldsheet connecting the two loops found in \cite{Correa:2018lyl} has been rewritten in terms of new parameters, to easily make a comparison between different saddle-points in the defect theory. We found that, if the first loop is placed at a non-zero distance from the defect and the value of $h/R$ is small, the dominant solution is the connected one. Increasing the separation distance between the loops, the other conceivable configurations mentioned above can become dominant, depending on the values of the other parameters of the system. We have also examined the case concerning the "string-brane crossing", which contemplates the possibility for the connected solution between the circles to touch and then intersect the defect brane. In this case, the transition between the connected solution and two separate cylindrical surfaces attached to the defect can happen before the latter configuration has a smaller area than the former, resulting in a zero-order phase transition. In \cite{Bonansea:2019rxh}, it was shown that, for a single circle, the dome geometry becomes less energetically preferred with respect to the cylindrical surface before touching the defect, i.e. the GO-like transition always happens before the string-brane crossing. In our case, since we have a larger parameter space, the discussion is more involved. We avoid this phenomenon by placing the string endpoints in the same hemisphere of the $S^5$.
 In principle, non-trivial string three-point functions could also enter the game, describing new connected minimal surfaces with three holes, one of which lying on the defect. This configuration involves string interactions and has an additional power of $1/N$ in the gauge theory language. Thus, it is a higher order effect respect to the classical string solutions that we considered in this work. Moreover, we expect this configuration's analysis to be very involved and we leave the study of its properties to future work. The two circles correlator's study in a defect version of $\mathcal{N}=4$ SYM can be extended to the case of circular Wilson loops with different radii, obviously increasing the dimensions of the parameter space. As pointed out in \cite{Drukker2005a}, the correlator of two circles in parallel planes separated by a distance $h$ are conformally related by a
stereographic projection to coincident circles with different radii. When the defect is present, the conformal map should also involve the defect itself, which spoils part of the conformal invariance of the original theory. As a future development, it would be interesting to analyze the effect of the stereographic projection on the defect and how this conformal map works in the defect CFT (dCFT) case. Moreover, by virtue of the enticing results achieved in \cite{Komatsu:2020sup}, it would be worthwhile to compute in the dCFT the expectation value of a circular Wilson loop operator and the two-circles correlator, in the BPS case, using the localization technique.
Also, the Wilson loops correlator in higher representations, in particular the symmetric and antisymmetric ones \cite{Tai:2006bt, Aguilera-Damia:2017znn}, can be inspected in this new defect set-up.

\subsection*{Acknowledgements}
Special thanks go to Luca Griguolo, Domenico Seminara, and Diego Trancanelli for participating in the early stages of this work and many interesting discussions. It is also our pleasure to thank D. Correa and C. Kristjansen, for useful discussions. The work of S.B. is supported by " Fondazione Angelo Della Riccia" and in part by the research fellowship "Non-perturbative methods in quantum field theory: holography and beyond" at Università degli Studi di Firenze, R.S was financed in part by the Coordena\c{c}\~ao de Aperfei\c{c}oamento de Pessoal de N\'ivel Superior - Brasil (CAPES) - Finance Code 001. 

\appendix
\section{Jacobi elliptic functions and elliptic integrals}
\label{elliptics}
In this paper we work with the standard Jacobi elliptic functions and elliptic integrals defined along the lines of \cite{Abram1964}. We use the incomplete elliptic integrals
\begin{align}
\elF{x}{m}&=\int_0^x\frac{d\th}{\sqrt{1-m\sin^2\th}}\;,  \noindent \\
\elE{x}{m}&=\int_0^x d\th\sqrt{1-m\sin^2\th}\;,\noindent \\
\Pi\left(n;x \left| m \right.\right) &=\int_0^x \frac{d\th}{\left(1-n\sin^2\th\right)\sqrt{1-m\sin^2\th}}.
\end{align}
of the first, second and third kind, respectively. The complete elliptic integrals are defined as
\be
\elF{\frac{\pi}{2}}{m}=\eK(m)\;, \quad \elE{\frac{\pi}{2}}{m}=\eE(m)\;, \quad \Pi\left(n; \frac{\pi}{2}\left| m \right.\right)=\eP{n}{m}\;.
\ee 
We also use the Jacobi amplitude $\vf=\text{am}(x|m)$ which is the inverse of $\elF{x}{m}$
\be
x=\elF{\text{am}(x|m)}{m}\;.
\ee
The Jacobi elliptic functions are defined as
\begin{equation}
\label{Jels}
\sn{x}{m}=\sin\vf,\quad\cn{x}{m}=\cos\vf\quad\text{and}\quad\dn{x}{m}=\sqrt{1-m\sin^2\vf},
\end{equation}
such that $\sn{\mathds{K}(m)}{m}=1$ and $\cn{\mathds{K}(m)}{m}=0$. 
 The reciprocals of the latter functions are
\begin{equation}
\label{Jiels}
\ns{x}{m}=\frac{1}{\sn{x}{m}},\quad\nc{x}{m}=\frac{1}{\cn{x}{m}},\quad\nd{x}{m}=\frac{1}{\dn{x}{m}}.
\end{equation}

  \section{Parametrization of the connected solution}
  We check that our parametrization reproduces the results of \cite{Correa:2018lyl} where the classical connected solution is written down in terms of the two parameters $s$ and $t$:
  \begin{align}
  &\frac{1}{4} \log\left(\frac{a+h+b}{a-h-b}\, \frac{a-b}{a+b} \right) = F(s,t)=\nonumber \\ &=\frac{\sqrt{t}}{\sqrt{s(1+t)}}\left[ \mathds{K}\left(\frac{s+t}{1+t} \right)-(1-s)\Pi\left(s \left|\frac{s+t}{1+t} \right.\right)  \right],\\ 
  &\label{eq:J.31}\g \equiv G(s,t)=2\,\frac{\sqrt{1-s-t}}{\sqrt{1+t}}\,\mathds{K}\left(\frac{s+t}{1+t} \right),\\
  &\label{eq:J.32}S^{reg}(s,t)=-\frac{2 \sqrt{\l}}{\sqrt{s}}\frac{1}{\sqrt{1+t}}\,\left[(1+t)\mathds{E}\left(\frac{s+t}{1+t} \right)-(1-s)\,\mathds{K}\left(\frac{s+t}{1+t} \right) \right],
  \end{align}
  To verify that equation \eqref{eq:J.31} corresponds to our result for $\g$ given in \eqref{gamma}, we take the definition of $s$ and $t$
  \be
  \label{eq:J.34}
  s=\frac{\sqrt{(1+K_{\theta}^2)^2+4\,a^2K_x^2}-(1+K_{\theta}^2)}{2\,a^2 K_x^2}, \qquad \qquad t=a^2K_x^2s^2\,,
  \ee
  where $K_x$ and $K_{\theta}$ are the two constants of motion found in \cite{Correa:2018lyl}. It is easy to check that they correspond exactly to $c$ and $j$, which are the positive integration constants given in \cite{Bonansea:2019rxh}
  \be
  K_x=-c=-\frac{1}{R}\sqrt{\frac{-(j^2+m) (j^2 m+1)}{(m+1)^2}}
  \text{sech}\left(\frac{v(\hat\s)}{2}\right)\,, \qquad K_{\theta}=j\,.
  \ee
  The expression for $a$, for the equal radii case, in our notation is 
  \be
  a=\sqrt{\frac{h^2}{4}+R^2}=R \cosh\left(\frac{v(\hat \s)}{2} \right) \,.
  \ee
  Therefore, we find that $s$ and $t$ can be written in terms of $j^2$ and $m$ in the following way
  \be
  \label{sandt}
  s=\frac{1+m}{j^2+m}\,, \qquad t=-
  \frac{1+m\,j^2}{j^2+m}\,.
  \ee
  If we rephrase the bounds on $s$ and $t$ 
  \be
  0\leq s\leq 1\,,\qquad 0 \leq t \leq 1-s
  \ee
  in terms of $m$ and $x=\sqrt{\frac{j^2-1}{j^2(m+1)}}$, we get exactly the same bounds on $x$ given in \eqref{boundsx}.
  If we perform the substitution of \eqref{sandt} in \eqref{eq:J.31} using the identity
  \be
  \mathds{K}\left(\frac{m}{m-1} \right) =\sqrt{1-m}\;\mathds{K}(m)\,,
  \ee
  we get for $\g$ the expression given in \eqref{gamma}.
  The same can be done for the action and substituting \eqref{sandt} in equation \eqref{eq:J.32} we get exactly \eqref{action} using the identity
  \be
  \mathds{E}\left(\frac{m}{m-1} \right) =\frac{1}{\sqrt{1-m}}\;\mathds{E}(m).
  \ee
  In order to verify that our solution reproduces also the form for $F(s,t)$, we notice that 
  \be
  \frac{1}{4} \log\left(\frac{a+h+b}{a-h-b}\, \frac{a-b}{a+b} \right) =\frac{v(\hat\s)}{2}\,,
  \ee
  where $b=-h/2$ in equal radii case. 
  The l.h.s of the equation above is equal to the integral
  \be
  \label{F}
  \int_{0}^{\theta_0} d \theta\, \frac{\sqrt{K_x^2\,a^2\,\sin^4 \theta}}{\sqrt{\cos^2\theta-K^2_\theta\,\sin^2 \theta-K^2_x\, a^2\sin^4 \theta}}
  \ee
  that, through the change of variable in \eqref{eq:J.34}, defines $F(s,t)$. This integral is obtained parametrizing $r$ and $y$ as
  \be
  \label{randy}
  r=\sqrt{a^2-(x_3+\tilde{b})^2}\cos \theta \qquad \text{and}\qquad y=\sqrt{a^2-(x_3+\tilde{b})^2}\sin \theta\,,
  \ee
  where $\tilde{b}=-h/2-L$. Since for us
  \be
  \sqrt{a^2-(x_3+\tilde{b})^2}=R\, \text{cosh}\left(\frac{v(\hat \s)}{2} \right)  \,\text{sech}\left( v(\s)-\frac{v(\hat \s)}{2}\right) \,,
  \ee
  if we compare \eqref{randy} with \eqref{solution}, where $\eta=v(\hat \s)/2$, we recognize that
  \be
  \label{change}
  \cos \theta=\frac{g(\s)}{\sqrt{1+g(\s)^2}} \qquad \text{and}\qquad \sin \theta=\frac{1}{\sqrt{1+g(\s)^2}}\,.
  \ee
  Thus,
  \be
  d \theta=-\frac{g'(\s)}{1+g(\s)^2}\,d \s
  \ee
  and $\theta$ is zero at $\s=0$ or $\s=\hat{\s}$. The maximum value of $\theta$, denoted as $\theta_0$, is reached when $g(\s)$ assumes its minimal, namely at
  \be
  \s_0=\frac{\mathds{K}(m)}{\sqrt{n}}=\frac{\hat\s}{2}\,.
  \ee
  Performing the change of variables \eqref{change} in \eqref{F}, we get
  \be
  \begin{split}
  	& \int_{0}^{\theta_0} d \theta\, \frac{\sqrt{K_x^2\,a^2\,\sin^4 \theta}}{\sqrt{\cos^2\theta-K^2_\theta\,\sin^2 \theta-K^2_x\, a^2\sin^4 \theta}}= \\ & \int_{0}^{\s_0}d \s\,\frac{-g'(\s)}{1+g(\s)^2}\,\frac{c\, a}{1+g(\s)^2}\frac{1+g(\s)^2}{\sqrt{g(\s)^4+(1-j^2)g(\s)^2+m\,\frac{(j^2-1)^2}{(m+1)^2}}}\,.
  \end{split}
  \ee
  Using the expression for the first integral for $g(\s)$ derived in \cite{Bonansea:2019rxh}
  \be
  g'(\s)^2=g(\s)^4+(1-j^2)g(\s)^2+m\,\frac{(j^2-1)^2}{(m+1)^2}
  \ee
  and the fact that $g'(\s)$ is negative if $0\leq \s \leq \s_0$, the previous integral simplifies to
  \be
  \sqrt{\e_0}\int_{0}^{\s_0}d \s\,\,\frac{-g'(\s)}{1+g(\s)^2}\,\frac{1}{-g'(\s)}=\sqrt{\e_0}\int_{0}^{\s_0}d \s\,\,\frac{1}{1+g(\s)^2}=v(\s_0)\,.
  \ee
  Performing the integral, we get
  \be
  \begin{split}
  	&v(\s_0)=\sqrt{\e_0}\left.\left(\s-\frac{1}{\sqrt{n}}\Pi\left(-\frac{1}{n};\text{am}\left(\sqrt{n}\,\s\left. \right| m \right)\left. \right| m  \right)  \right) \right| ^{\s_0}_0=\\ &=\sqrt{\frac{\e_0}{n}}\left(\mathds{K}(m) -\Pi\left(-\frac{1}{n} \left. \right| m \right) \right) =\frac{v(\hat\s)}{2} \,.
  \end{split}
  \ee
  Finally, we have shown that also $F(s,t)$ can be perfectly mapped to our parametrization.
  \section{Explicit solution for the string-brane crossing}
  \label{appB}
  We can use a combination of conditions  \eqref{condder} and \eqref{cond1} to express $\kappa_*$ as 
  \be
  \k_*=\frac{\sqrt{\e_0} \sqrt{g(\s_*)^2+1}}{\sqrt{\e_0} \sinh \left( v(\hat{\s})/2-v(\s_*)\right) -g(\s_*) g'(\s_*) \cosh\left(  v(\hat{\s})/2-v(\s_*)\right) }\;,
  \ee
  where we have also used that
  \be
 x'_3(\s) =\frac{\sqrt{\e_0}}{R}\sech\left(\frac{v(\hat{\s})}{2} \right) y^2(\s)\;.
  \ee
  Notice that $\e_0$, defined in eq. \eqref{eps0}, can be expressed as a function of $m$ and $\gamma$
  \be
\e_0= \frac{-\left(\g ^2-4 \mathds{K}(m)^2\right) \left(\g ^2-4 m\mathds{K}(m)^2\right)}{\left(\g ^2-4 (m+1)
  		\mathds{K}(m)^2\right)^2}\;.
  	\ee
  The critical value of $L/R$ is derived from eq. \eqref{cond1} using the explicit form of $x_3$ and $y$ given in eq. \eqref{solution} and it reads
  \be
  \left( \frac{L}{R}\right)_*=\frac{\sqrt{\e_0} \cosh
  	(v(\s_*))+g(\s_*) g'(\s_*) \sinh (v(\s_*))}{\sqrt{\e_0} \sinh \left( v(\hat{\s})/2-v(\s_*)\right) -g(\s_*) g'(\s_*) \cosh\left(  v(\hat{\s})/2-v(\s_*)\right) }\,.
  \ee
  \bibliographystyle{elsarticle-num}
  \bibliography{Mybib}
\end{document}